\documentclass[prd,superscriptaddress,nofootinbib]{revtex4}

\usepackage{amsmath}
\usepackage{amsfonts}

\usepackage{graphicx}
\usepackage{bm}

\def\beq{\begin{equation}}   
\def\eeq{\end{equation}}
\def\bed{\begin{description}}
\def\eed{\end{description}}
\def\susy{{\sc Susy}}

\def\bea{\begin{eqnarray}}  
\def\eea{\end{eqnarray}}
\def\wt{\widetilde}
\def\b{\beta}

\def\k{\kappa}
\def\la{\lambda}
\def\tb{\tan \beta}

\def\q{\quad}
\def\half{\frac{1}{2}\,}

\def\nmssm{NMSSM}
\def\mssm{MSSM}
\def\s{\hat{S}}  
\def\nmhdecay{{\tt {\sc Nmhdecay}}}
\def\micro{{\tt {\sc micrOMEGAs}}}
\def\lsp{{\widetilde{\chi}_1^0}}
\def\gev{{\rm GeV}}
\def\noi{\noindent}
\def\ba{\begin{array}}
\def\ea{\end{array}}
\def\nn{\nonumber}
\def\ds{{\tt {\sc DarkSUSY}}}

\begin{document}
\title{Indirect detection of light neutralino dark matter in the NMSSM}
\author{Francesc Ferrer}
\affiliation{
CERCA, Department of Physics, Case Western Reserve University,
10900 Euclid Avenue, Cleveland, OH 44106-7079, USA.}
\author{Lawrence M. Krauss}
\affiliation{
CERCA, Department of Physics, Case Western Reserve University,
10900 Euclid Avenue, Cleveland, OH 44106-7079, USA.}
\affiliation{ also Department of Astronomy, CWRU}
\author{Stefano Profumo}
\affiliation{California Institute of Technology, Mail Code 106-38, Pasadena, CA 91125, USA}

\begin{abstract}
We explore the prospects for indirect detection of neutralino dark matter in supersymmetric models with an extended Higgs sector (NMSSM). We compute, for the first time, one-loop amplitudes for NMSSM neutralino pair annihilation into two photons and two gluons, and point out that extra diagrams (with respect to the MSSM), featuring a potentially light CP-odd Higgs boson exchange, can strongly enhance these radiative modes. Expected signals in neutrino telescopes due to the annihilation of relic neutralinos in the Sun and in the Earth are evaluated, as well as the prospects of detection of a neutralino annihilation signal in space-based gamma-ray, antiproton and positron search experiments, and at low-energy antideuteron searches. We find that in the low mass regime the signals from capture in the Earth are enhanced compared to the MSSM, and that NMSSM neutralinos have a remote possibility of affecting solar dynamics.  Also, antimatter experiments are an excellent probe of galactic NMSSM dark matter. We also find enhanced two photon decay modes that make the possibility of the detection of a monochromatic gamma-ray line within the NMSSM more promising than in the MSSM. 

\end{abstract}

\date{September 25, 2006}
\maketitle

\section{Introduction}

Numerous theoretical and phenomenological motivations exist for a Minimal Supersymmetric extension of the Standard Model  (MSSM).  At the same time one of the attractive by-products of low energy supersymmetry is the natural occurrence in the particle content of the theory of a stable weakly interacting massive particle, the lightest neutralino, which could be the microscopic constituent of the as yet unobserved galactic halo dark matter. Another strong motivation comes from the SM hierarchy problem, originating from the large fine-tuning required by the stability of the electroweak scale to radiative corrections, originating from the large number of orders of magnitude occurring between the GUT, or the Planck, scale, and the electroweak scale itself.

Although very appealing, the MSSM has been challenged by various pieces of experimental information, and by some arguments of more theoretical nature. Among these, the LEP-II limit on the mass of the lightest CP-even Higgs~\cite{Bastero-Gil:2000bw}, the constraints on the masses of supersymmetric (\susy) charged or colored particles from direct searches at LEP and at the Tevatron \cite{pdg}, and the so-called $\mu$ problem, {\em i.e.} the fundamental reason why the \susy\ Higgsino mass term $\mu$ appearing in the MSSM superpotential lies at some scale near the electroweak scale rather than at some much higher scale.

The addition of a new gauge singlet chiral multiplet, $\s$,
to the particle content of the MSSM can provide an elegant solution 
to the mentioned $\mu$ problem of the MSSM~\cite{Kim:1983dt}. The so-called Next to Minimal
Supersymmetric Standard Model (NMSSM)~\cite{Ellis:1988er}
is an example of one such minimal
extension that also alleviates the {\it little fine tuning problem} of the
\mssm, arising from the non-detection of a neutral CP-even Higgs at 
LEP-II~\cite{Bastero-Gil:2000bw}. 

A further motivation to go beyond the MSSM comes from Electro-Weak Baryogenesis (EWB), {\em i.e.} the possibility that the baryon asymmetry of the Universe originated through electro-weak physics at the electro-weak phase-transition in the Early Universe. Although still a viable scenario within the MSSM \cite{ewb}, EWB generically requires the Higgs mass to be in the narrow mass range above the current LEP-II limits and below $\simeq 120$ GeV, a rather unnatural mass splitting between the right-handed and the left-handed stops (the first one required to lie below the top quark mass, and the other in the multi-TeV range), CP violation at levels sometimes at odds with electric dipole moment experimental results, and, generically, a very heavy sfermion sector \cite{ewbmssm}. In contrast, the NMSSM provides extra triscalar Higgs couplings which hugely facilitate the occurrence of a more strongly first-order EW phase transition, and extra CP violating sources, relaxing most of the above mentioned requirements in the context of the MSSM~\cite{ewbnmssm,Funakubo:2005pu}.

%Furthermore, in contrast with the \mssm,
%electroweak baryogenesis remains a viable possibility 
%within the \nmssm~\cite{Funakubo:2005pu}.

One of the chief remaining cosmological issues associated with the NMSSM,  the cosmological 
domain wall problem~\cite{Abel:1995wk}, 
caused by the discrete $\mathbb{Z}_3$ symmetry of the \nmssm,
can be circumvented by introducing non-renormalizable Planck-suppressed 
operators~\cite{Panagiotakopoulos:1998yw}.

The Higgs sector of the \nmssm\ contains three CP-even and two CP-odd scalars,
which are mixtures of \mssm-like Higgses and singlets. Also,
the neutralino sector contains five mass-eigenstates, instead of the four in
the \mssm, each of which has, in addition to the four \mssm\ components, a
singlino component, the latter being the fermionic partner of the extra singlet scalars. The extended Higgs and neutralino sectors weaken the mass bounds for 
both the Higgs bosons and the neutralinos. Very light neutralinos
and Higgs bosons, even in the few GeV range, are in fact not excluded in the \nmssm~\cite{Franke:1994hj} (see also~\cite{Ellis:1988er,Franke:1995tc}; the particle spectrum with the dominant 1-loop and 2-loop corrections to the Higgs sector is available via the numerical code \nmhdecay~\cite{Ellwanger:2005dv}).

The cosmology of Dark Matter singlinos has been addressed long ago in 
Ref.~\cite{Greene:1986th}, while the LSP relic abundance in the \nmssm\ was
first calculated in~\cite{Abel:1992ts}. Constraints from electroweak
symmetry breaking and GUT scale universality were added 
in~\cite{Stephan:1997ds}. More recently, the computer code
\micro~\cite{Belanger:2006is}, has been extended to allow for the relic
density calculation in the \nmssm~\cite{Belanger:2005kh}\footnote{
The phenomenology of the lightest neutralino in a different extension of the \mssm, the Left-Right \susy\ model, has been recently surveyed in \cite{Demir:2006ef}.}. %The physics of light singlinos has recently scrutinized in Ref.~\cite{Gunion:2005rw}

The implications for the direct detection of \nmssm\ neutralinos were first
studied in~\cite{Bednyakov:1998is}, where light neutralinos ($\sim$ 3 GeV) 
with acceptable relic abundance and sufficiently large expected event rates
for direct detection with a ${}^{73}$Ge-detector were found in different
domains of the parameter space, when the gaugino unification 
relation~\cite{Stephan:1997ds} was relaxed. In~\cite{Cerdeno:2004xw},
the theoretical predictions for the spin-independent neutralino-proton
cross section, $\sigma_{\lsp-p}$, 
were reevaluated, and all available experimental constraints
from LEP on the parameter space were taken into account. Values within reach
of present dark matter detectors were obtained in regions with 
very light Higgses, $m_{h_1^0} \lesssim 70 \gev$, with a significant singlet
contribution. The lightest neutralino, in those regions, features a large 
singlino-Higgsino composition, and a a mass in the range 
$50 \gev \lesssim m_\lsp \lesssim 100 \gev$. More recently, \nmssm\
neutralinos as light
as $100 {\rm MeV} \lesssim m_\lsp \lesssim 20 \gev$, satisfying accelerator
constraints and with the right relic density, have been shown
to occur in~\cite{Gunion:2005rw}, where it was argued that the
\nmssm\ can, moreover, provide neutralinos in the mass range that would
be required to reconcile the DAMA claim of discovery with the limits
placed by CDMS.

So far, theoretical studies have not addressed the possibility that \nmssm\ neutralinos making up the galactic dark matter can manifest themselves {\em indirectly}. For instance, pair annihilations of neutralinos in the galactic halo can produce sizable amounts of antimatter, which current and forthcoming space-based antimatter search experiments can possibly detect; neutralinos trapped in the Sun or in the Earth~\cite{Krauss:1985ks, Krauss:1985aa} can give rise to a coherent flux of energetic neutrinos from the center of the Sun or of the Earth; pair annihilation of neutralinos, either in nearby large-dark-matter-density sites, or from the cumulative effect of annihilations outside the Galaxy, can produce gamma-ray fluxes at a level detectable by GLAST or by ground-based air Cherenkov Telescopes; and last, but not least, the exciting possibility of peculiar gamma-ray spectral features, like a sharp monochromatic peak at $E_\gamma\simeq m_\chi$ (where $\chi$ indicates the lightest neutralino, assumed to be the lightest supersymmetric particle), from loop-induced $\chi\chi\rightarrow\gamma\gamma$ processes, can also be, in principle, very promising.

A first motivation for looking into indirect dark matter detection within the \nmssm\ comes from the possibility that thermally produced neutralinos, in this context, can be {\em very light}. Since the pair annihilation rate of thermal relics is roughly fixed by requiring that the thermal neutralino abundance coincides with the cold dark matter abundance inferred by astrophysical observations \cite{Spergel:2006hy}, the indirect detection rates generically scale as $1/m_\chi^2$: light neutralinos are therefore expected to give significantly enhanced rates with respect to the standard case.

A more technical point has provided us with a second motivation to look into indirect detection prospects for \nmssm\ neutralinos: loop induced pair annihilations of neutralinos into two photons or two gluons (respectively contributing to the mentioned monochromatic gamma-ray line and to, {\em e.g.}, antimatter fluxes) are predicted to be increased, within the \nmssm, by the presence of extra diagrams mediated by the (potentially light) extra CP-odd Higgs boson. We therefore extend here, for the first time, the MSSM results for these loop-induced neutralino pair-annihilation amplitudes \cite{Bergstrom:1997fh} to the NMSSM.

Our results suggest several signatures, including muons resulting from neutralino annihilation in the Earth, and antiparticle and gamma ray production from neutralino annihilation in the galaxy, where the NMSSM produces signals that are enhanced compared to those predicted in the MSSM. 

The outline of this article is as follows: we  first introduce the theoretical framework and set our notation in Sec.~\ref{sec:nmssm}; we devote Sec.~\ref{sec:paramspace} to a discussion of the viable \nmssm\ parameter space. Sec.~\ref{sec:indirectdet} contains our central results on indirect \nmssm\ neutralino dark matter detection, while the appendices provide the reader with details on the relevant \nmssm\ neutralino pair-annihilation amplitudes and on neutralino-nucleon scattering cross sections.

%Apart from the direct detection of Dark Matter by monitoring the elastic
%scattering on target nuclei, 
%stringent limits for \mssm-like neutralinos are provided by
%indirect methods related to the different annihilation products of neutralinos in the halo of the Galaxy or those trapped in the Sun or in the Earth. In this
%paper, we study the indirect signals from \nmssm-like
%neutralinos and the constraints that can be placed by present or future
%earth-based or satellite experiments. {\bf FF: Add \mssm references here}.

\section{The NMSSM}\label{sec:nmssm}

We hereby describe the Lagrangian of the \nmssm. Our notation follows
that of the code \nmhdecay~\cite{Ellwanger:2005dv}, which we have used to
explore the \nmssm\ parameter space\footnote{Note that the Higgs states
$H_u$, $H_d$ are usually denoted in the \mssm\ by $H_2$ and $H_1$. As shown in
the Appendix, some indices in both the neutralino and the Higgs mass
matrices need to be switched accordingly to make contact with the corresponding
\mssm\ expressions.}.

Apart from the usual quark and lepton Yukawa couplings, 
the scale invariant superpotential is\footnote{Hatted capital 
letters denote superfields, and unhatted ones their scalar components.}
\beq 
\lambda \ \widehat{S} \widehat{H}_u
\widehat{H}_d + \frac{\kappa}{3} \ \widehat{S}^3 
\label{superpot}
\eeq
\noindent depending on two dimensionless couplings 
$\lambda$, $\kappa$ beyond the MSSM, and the 
associated trilinear soft-\susy-breaking terms
\beq 
\label{trilinear} 
\lambda A_{\lambda} S H_u H_d + \frac{\kappa}
{3} A_\kappa S^3\,. 
\eeq
The two other input parameters, 
$\tan \beta =\ \left< H_u \right>/ \left< H_d \right>$ and
$\mu_\mathrm{eff} = \lambda \left< S \right>$, 
along with $M_Z$, determine the
three \susy\ breaking masses squared for $H_u$, $H_d$
and $S$ through the three minimization equations of the
scalar potential. Note that an effective $\mu$-term is generated
from the first term in Eq.~(\ref{superpot}) for a non-zero value
of the VEV $\left< S \right>$. With the sign conventions 
of~\cite{Ellwanger:2005dv} for the fields, $\lambda$ and $\tan \beta$
are positive, while $\k$, $A_{\lambda}$, $A_{\k}$ and $\mu_\mathrm{eff}$
can have either sign.

Assuming CP conservation in the Higgs sector, there is no mixing between
CP-even and CP-odd Higgses.
More concretely, for VEVs $h_u \equiv \left< H_u \right>$,
$h_d \equiv \left< H_d \right>$ and $s \equiv \left< S \right>$ such that
\beq
H_u^0 = h_u + \frac{H_{uR} + iH_{uI}}{\sqrt{2}} , \q
H_d^0 = h_d + \frac{H_{dR} + iH_{dI}}{\sqrt{2}} , \q
S = s + \frac{S_R + iS_I}{\sqrt{2}},
\eeq
the CP-even mass matrix in the basis $S^{bare} = (H_{uR}, H_{dR}, S_R)$ is
rendered diagonal by an orthogonal matrix $S_{ij}$. One, thus, obtains
3 CP-even mass eigenstates $h_i=S_{ij} S^{bare}_j$, with increasing masses
$m_{h_i}$. The bare CP-odd states $P^{bare} = (H_{uI}, H_{dI}, S_I)$ are
related to the physical CP-odd states $a_i$, $i=1,2$, and the
massless Goldstone mode $a_3\equiv \wt{G}$ by $a_i=P_{ij} P^{bare}_j$, where
$a_1$ and $a_2$ are ordered with increasing mass. Details of the bare mass
matrices in terms of the \nmssm\ parameters can be found 
in~\cite{Ellwanger:2005dv}.

With fixed parameters of the Higgs sector, the masses and mixing of the
neutralinos are determined by two additional parameters: the masses $M_1$
and $M_2$ of the $U(1)_Y$ gaugino, $\la_1$, and the neutral $SU(2)$ gaugino,
$\la_2^3$. In the basis $\psi^0 = (-i\la_1 , -i\la_2, \psi_u^0, \psi_d^0,
\psi_s)$ the symmetric mass matrix ${\cal M}_0$ of the neutralinos
\beq
{\cal L} = - \half (\psi^0)^T {\cal M}_0 (\psi^0) + \mathrm{h.c.},
\eeq

\noi has the form
\beq
{\cal M}_0 =
\left( \ba{ccccc}
M_1 & 0 & \frac{g_1 h_u}{\sqrt{2}} & -\frac{g_1 h_d}{\sqrt{2}} & 0 \\
& M_2 & -\frac{g_2 h_u}{\sqrt{2}} & \frac{g_2 h_d}{\sqrt{2}} & 0 \\
& & 0 & -\mu & -\la h_d \\
& & & 0 & -\la h_u \\
& & & & 2 \k s
\label{neutralinomass}
\ea \right) . \eeq

This matrix can be diagonalized by a real orthogonal matrix, $N_{ij}$,
obtaining 5 eigenstates, $\chi^0_i = N_{ij} \psi^0_j$,  with real, but
not necessary positive masses, $m_\mathrm{\chi^0_i}$, ordered in increasing
absolute value of the mass\footnote{The matrix~(\ref{neutralinomass}) can
also be diagonalized using a complex $N_{ij}$. In that case the mass eigenstates
would be real and positive. These two choices result in different signs of
certain Feynman rules, as pointed out in the Appendix (for details
see~\cite{Gunion:1984yn}).}.

\subsection{Light neutralino dark matter: parameter space}\label{sec:paramspace}

Even though our study of the indirect detection of \nmssm-like neutralinos
is completely general, we choose to 
focus on light neutralinos, $m_\lsp \lesssim 100 \gev$, since 
the differences with the case of the \mssm\ will be more acute in this case.

We have performed a scan of the parameter space with the program \nmhdecay.
For each point, after computing the masses and couplings of all physical
states in the Higgs, chargino and neutralino sectors,
\nmhdecay\ checks for the absence of Landau singularities
below the GUT scale for $\la$, $\k$ and the Yukawa couplings $h_t$ and $h_b$.
This translates into  $\la < .75$, $\k < .65$, and
$1.7 < \tb < 54$~\cite{Belanger:2005kh}. \nmhdecay\ also checks for the absence 
of an unphysical global minimum of the scalar potential 
with vanishing Higgs VEVs\footnote{Tree level restrictions in parameter
space leading to valid minima are discussed in~\cite{Cerdeno:2004xw}.}. 
The program also makes sure that Higgs and squark
masses are positive, thus avoiding, in particular, charge breaking minima.

Finally, the available experimental constraints from LEP are imposed, including
unconventional channels relevant for the \nmssm\ Higgs sector, bounds
on the invisible $Z$ width (for light neutralinos) and limits on
chargino and neutralino pair production.

As remarked in~\cite{Cerdeno:2004xw}, there are other experimental bounds
that might put constraints on the parameter space. Rare B-meson decays,
sensitive to physics beyond the Standard Model like supersymmetry, have 
been studied for the \nmssm\ in the large $\tan \beta$ 
regime~\cite{Hiller:2004ii}. However, the transitions $b \rightarrow s \gamma$
and $B_s \rightarrow \mu^+ \mu^-$ are both flavor changing while the
contributions from a light $\lsp$ can be suppressed by making the appropriate
squark or slepton heavy~\cite{Gunion:2005rw}. 

Additional constraints, that apply when the $\lsp$ and $a_1$ are light,
were considered in~\cite{Gunion:2005rw}. The conclusion is that bounds on 
the magnetic 
moment of the muon can only be violated in extreme models, while 
rare kaon decays rule out some models with extremely light $a_1$. 

On the other hand, decays of 
the vector resonances $J/\Psi$ and $\Upsilon$ might be important for
models with a light $\lsp$ and/or $a_1$ (we follow here
the discussion in~\cite{Gunion:2005rw}). 
In some of our models, the decay
$V \rightarrow a_1 \gamma$, where $V$ stands for $J/\Psi$ or $\Upsilon$, is
indeed possible\footnote{Our scan of the \nmssm\ parameter space does not
yield $\lsp$ light enough to make the decay $V\rightarrow \gamma \lsp \lsp$
kinematically allowed.}. The width, relative to the muon decay channel is
at leading order~\cite{Wilczek:1977pj}:
\beq
\frac{\Gamma \left( V \rightarrow \gamma a_1 \right)}
{\Gamma \left( V \rightarrow \mu \mu \right)}=
\frac{G_F m_b^2}{\sqrt{2} \alpha \pi}
\left(1-\frac{m_{a_1}^2}{m_V^2} \right) X^2,
\label{jpsibr}
\eeq
\noi where $X=\tan \beta P'_{11}$ for the $\Upsilon$ decay,
$X={\rm cotan} \beta P'_{11}$ for the $J/\Psi$ and the $P'_{11}$ gives the
piece of $a_1$ that would be the \mssm\ pseudoscalar if the singlet were
not present\footnote{The definition of $P'_{ij}$ in terms of the CP-odd
Higgs mixing matrix, $P_{ij}$, can be found in~\cite{Ellwanger:2005dv}.}.

The ratio in Eq.~(\ref{jpsibr}) is generally less that $4 \times 10^{-9}$
($0.006$) for $J/\Psi$ ($\Upsilon$) decays, below the
CLEO measurement of $\Upsilon \rightarrow {\rm invisible} + \gamma$
\cite{Balest:1994ch}, although the highest branching ratios could be
discovered with new upcoming data or reanalyzing the existing CLEO data
\footnote{We find one model in our scan, with $m_{a_1}=3.01\ \gev$ and
$\tan \beta = 48.24$, yielding a ratio for the $\Upsilon$ decay large
enough to be already excluded by CLEO.}. On the other hand, lepton universality
tests in $\Upsilon$ decays by high-luminosity B factories could 
detect a CP-odd Higgs in the mass range $5\ \gev \lesssim m_{a_1}
\lesssim m_\Upsilon$, which would be otherwise very hard to discover by just
looking at the $a_1 \gamma$ channel~\cite{Sanchis-Lozano:2005di}.

To generate our models, we scan, at random, 
 the \nmssm\  parameter space in the region:
\bea
0 \leq &\la& \leq 0.75 \nonumber \\
-0.65 \leq &\k& \leq 0.65 \nonumber \\
1.7 \leq &\tan \beta& \leq 54 \nonumber \\
80 \gev \leq &\mu& \leq 500 \gev\nonumber \\
-500 \leq &A_\la,A_\k & \leq 500.
\label{scan}
\eea
The gaugino masses were also randomly chosen within the
bounds $0\: \gev \leq M_1 \leq 100 \:\gev $, 
$M_1 \leq M_2 \leq 500\:\gev $ and 
$300\: \gev \leq M_3 \leq 1000\:\gev $. The soft sfermion masses were
set to $M_3$, and the sfermion trilinear terms were varied within $\pm 1.5M_3$.

The models that passed the phenomenological constraints imposed by \nmhdecay\ were fed into
\micro\ to calculate the $\lsp$ relic density, taking into account
all possible annihilation and coannihilation channels. We kept as viable those
models that fell within the $2-\sigma$ region for the Cold Dark Matter abundance inferred by the WMAP team for a $\Lambda$CDM cosmology~\cite{Spergel:2006hy}.

\begin{figure}[htb]
\begin{center}
\begin{tabular}{c@{\hspace{1cm}}c}
\includegraphics[width=0.45\textwidth]{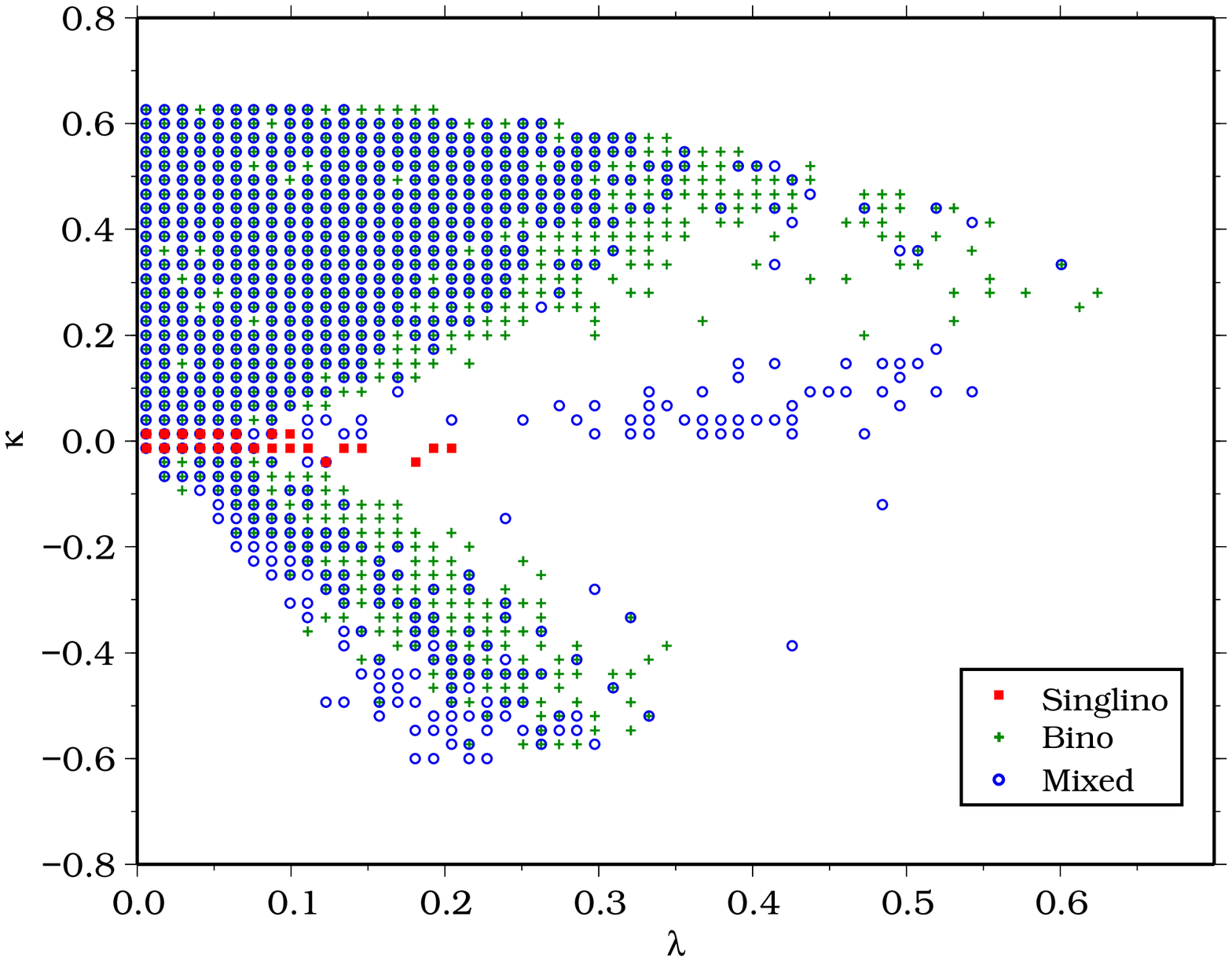}&
\includegraphics[width=0.45\textwidth]{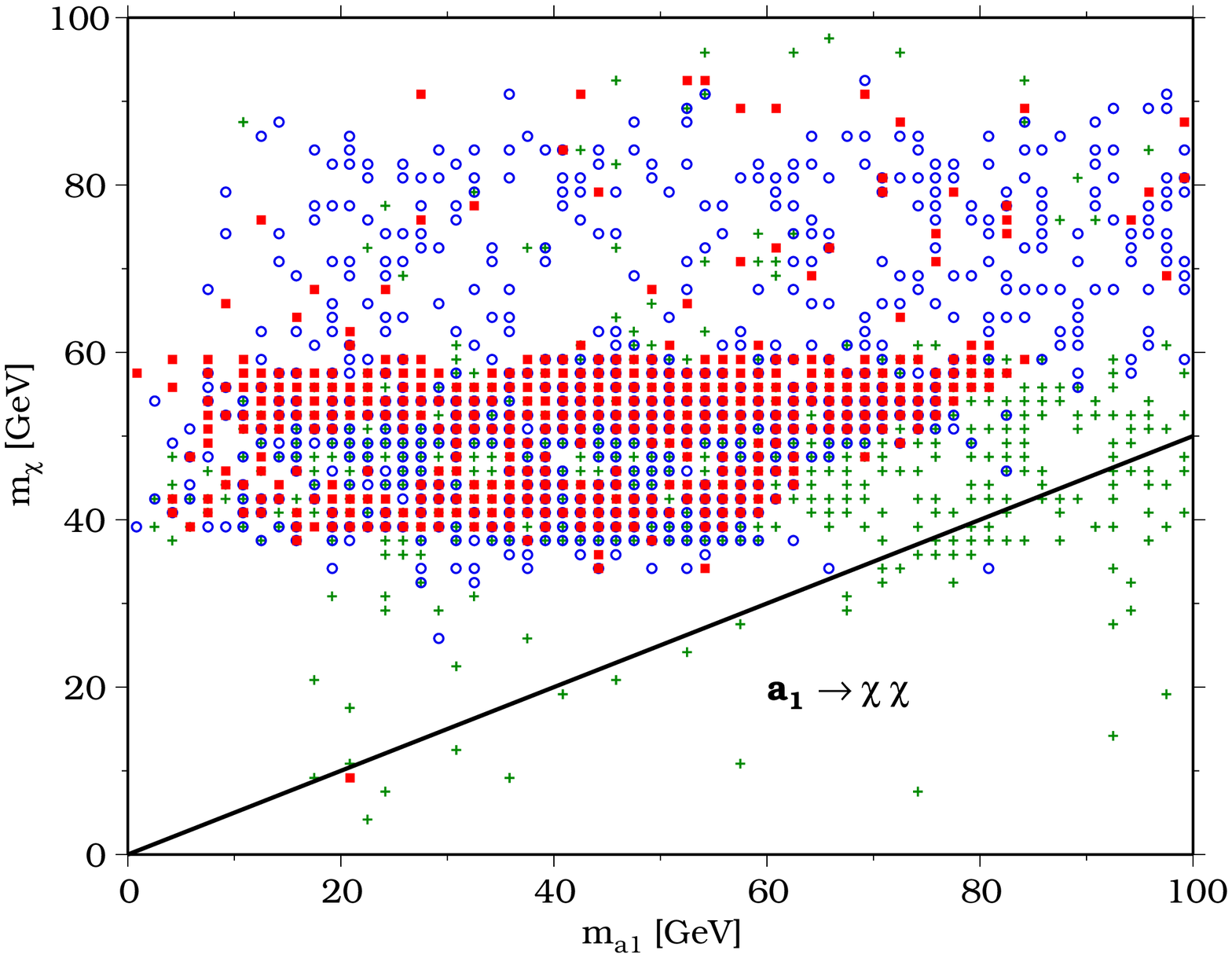}\\
& \\
\includegraphics[width=0.45\textwidth]{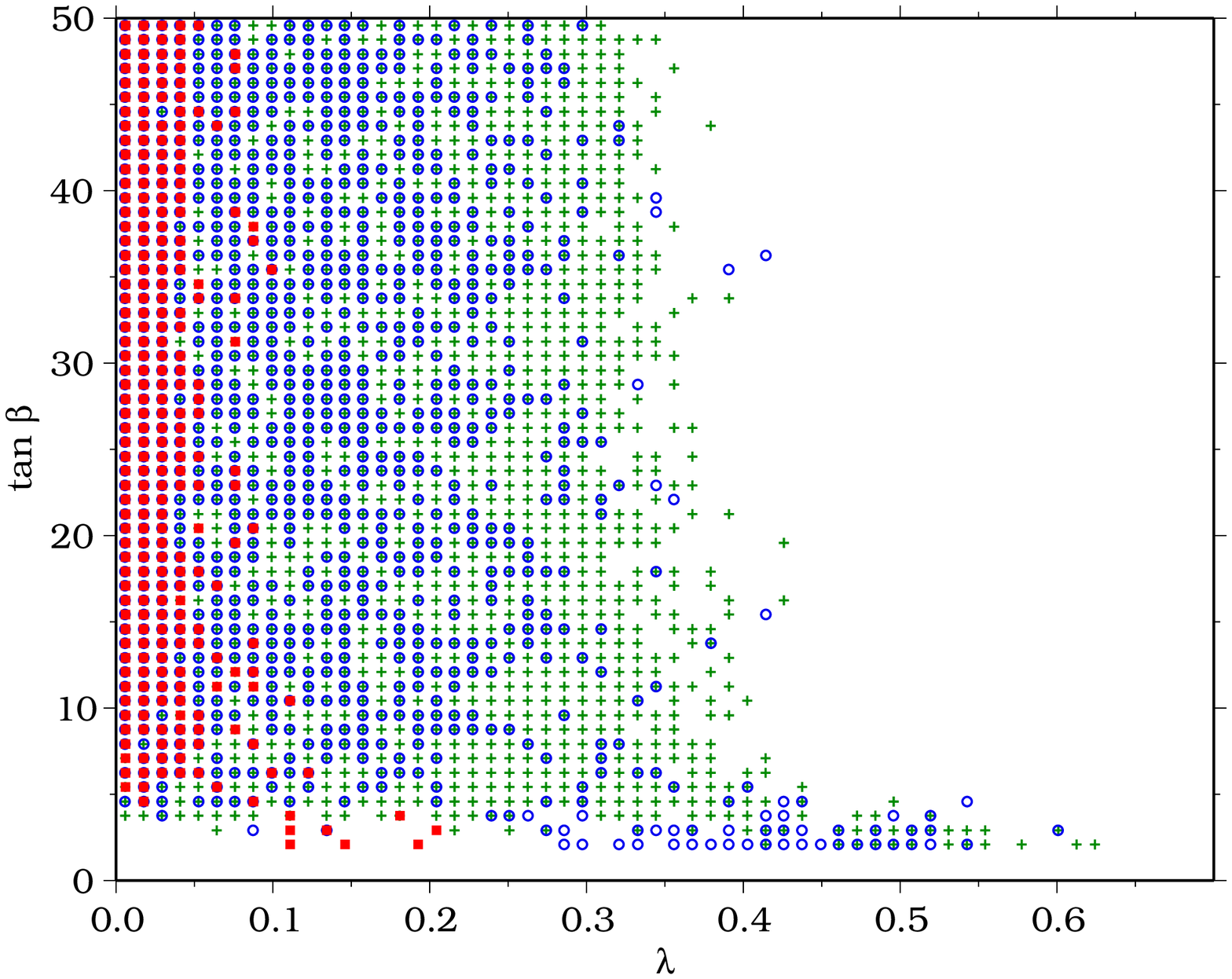}&
\includegraphics[width=0.45\textwidth]{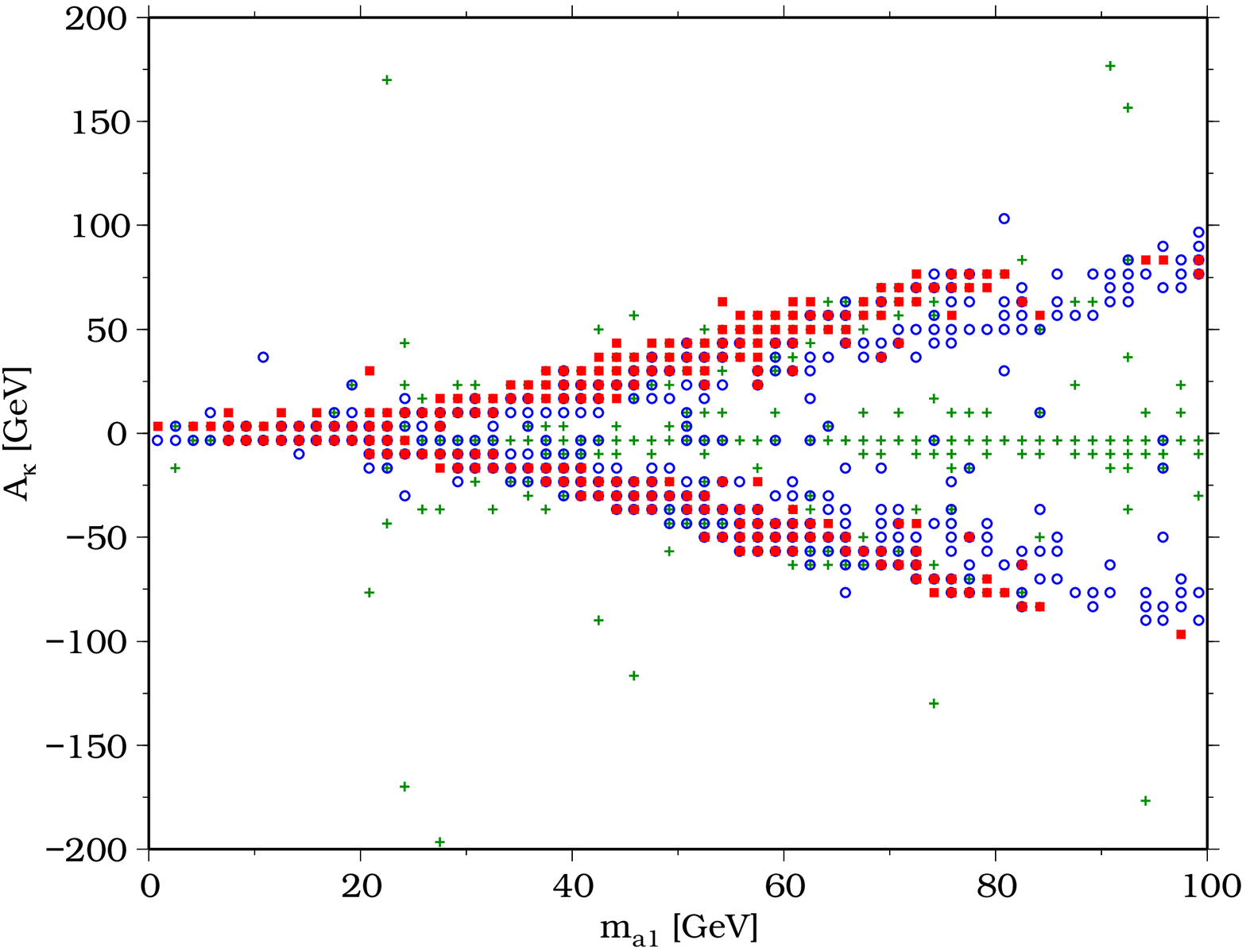}
\end{tabular}
\end{center}
\caption{The allowed region of $(\lambda-\kappa)$ parameter space (upper-left); 
the $m_\lsp$ vs. $m_{a_1}$ plane (upper-right)-- the decay $a_1 \rightarrow \lsp \lsp$
is allowed for those models below the line; the $(\tan \beta-\lambda)$ viable 
parameter space (lower-left); $A_\kappa$ as a function of the lightest
CP-odd Higgs mass (lower-right)}
\label{fignmssm}
\end{figure}

We show in Fig.~\ref{fignmssm} 
the region of the \nmssm\ parameter
space that satisfies the constraints discussed above. We have classified an
\nmssm-like neutralino as bino-like if $N_{11}^2 > 0.9$, singlino-like
if fulfilling the condition $N_{15}^2 > 0.9$, otherwise we indicate the neutralino ``mixed''.

As can be seen from Fig.~\ref{fignmssm}, the neutralino is mostly a singlino
when $\kappa$ and $\lambda$ are small. We can understand this feature
by realizing that the upper $4 \times 4$ block 
in the neutralino mass matrix, 
Eq.~(\ref{neutralinomass}), corresponds to the \mssm. From the lower
$3 \times 3$ block it can be appreciated that the singlino decouples from the \mssm\ part
when~\cite{Franke:2001nx}:
\beq
2 | \k s|, \: \la v < M_1,\: M_2, \: |\mu|.
\label{singlinoscan}
\eeq

It must be stressed that our Fig.~\ref{fignmssm} shows only models which give
an acceptable relic density. This might explain the absence of singlino-like
neutralinos at moderate $\lambda \sim 0.3$.

A CP-odd Higgs so light that the decay $a_1 \rightarrow \lsp \lsp$ is possible,
is generally not viable for light singlino-like neutralinos. This makes them
cosmologically disfavored, since this resonant decay is required to 
enhance the annihilation cross-section and obtain the correct relic density.
A nearly complete mass degeneracy between $\lsp$ and either the next-to-lightest neutralino, $\widetilde{\chi}_2^0$, or
the lightest chargino, suppressing the LSP final relic abundance through large coannihilation effects, usually occurs for the viable
light singlino models~\cite{Belanger:2006is}.

We can see from Fig.~\ref{fignmssm} that singlino models at large $\la$
feature small $\tan \beta \lesssim 5$, since large values of $\tan \beta$ induce
sizable singlino mixing. We also expect models with moderate $\tan \beta$,
for which annihilation through a Higgs
resonance is marginal in the \mssm, to be peculiar of the \nmssm\ setup.

Finally, the lower-right panel in Fig.~\ref{fignmssm} shows how a light
CP-odd Higgs boson, $a_1$, appears when $A_\kappa \rightarrow 0$. We stress that this
regime, when the $U(1)_R$ symmetry approximately holds, is well
motivated in the context of gaugino mediated \susy\ breaking \cite{Chacko:1999mi} where $A_\kappa$ is only
generated at the two loop level.

\section{NMSSM Dark Matter Indirect detection}\label{sec:indirectdet}
\label{sec:id}

Neutralinos, being weakly interacting and electrically neutral particles are very
difficult to observe in collider experiments directly. If they make
up a sizable fraction of the galactic halo dark matter, however, other methods
of detection become feasible~\cite{Bertone:2004pz,Jungman:1995df}.

Monitoring the energy deposited as they scatter off nuclei in detectors
falls into the realm of {\it direct detection} methods. A group of experiments
is actively exploring this path, although, as already mentioned, their
sensitivity decreases for neutralinos below 
$m_\chi \lesssim 100\ \gev$\footnote{
In this section we denote the lightest neutralino, $\lsp$, simply by $\chi$.}.
The prospects for direct detection of \nmssm\ neutralinos have already been
discussed in the 
literature~\cite{Bednyakov:1998is,Cerdeno:2004xw,Gunion:2005rw}.

We focus here on the possibility that dark matter neutralinos can be detected by
looking at products of their pair annihilation. Chief among them are neutrinos,
photons and antiparticles~\cite{Bertone:2004pz,Jungman:1995df,Feng:2000zu,
Zeldovich:1980st}. 

Neutrino fluxes from neutralino annihilations are searched for in
underground neutrino telescopes. Present facilities such as Super-Kamiokande
and MACRO, with low energy thresholds, $E_\nu \gtrsim 1\ \gev$, are
particularly useful to constrain the light \nmssm\ dark matter particles
that we consider. Some of the planned facilities ({\em e.g.} AMANDA, ICECUBE)
are geared to detecting neutrinos above $100\ \gev$, and we don't consider
them here. ANTARES, on the other hand, promises to improve the sensitivity
to moderately energetic neutrinos by an order of magnitude and will be of
importance for our discussion.

Gamma-rays are also produced in neutralino annihilations. They can be detected
by earth-based Cherenkov telescopes (MAGIC, HESS, VERITAS, \ldots) or in
space-borne facilities (EGRET, GLAST, AMS), although, 
only the latter have the ability
to observe the low energy photons from light neutralinos such as those considered here.

Other satellites, like PAMELA, GAPS and AMS, will measure the flux of antiparticles and
antimatter nuclei. 

We study below the signatures of light \nmssm\ dark matter particles in
neutrino telescopes, gamma-ray satellites and antimatter detectors. We also
touch upon the effects on the Sun caused by neutralino energy transport.

The main ingredient for indirect detection prediction are the different
annihilation modes of neutralinos. Since dark matter in the halo moves
at non-relativistic velocities, $v \sim 10^{-3} c$, only the channels
with a CP-odd final state can occur. The branching ratios for the
relevant tree level processes are reviewed in App.~\ref{appannihilation},
together with the most important one loop channels.

Neutrino fluxes from neutralino annihilations are enhanced in the direction of the center of the Sun or of the Earth. The abundance of neutralinos trapped within these objects depends on the scattering 
cross sections of neutralinos with nuclei that can be found in App.~\ref{appelastic}.

We start our discussion by studying the information that can be gained from the 
observation of neutrino fluxes.

\subsection{Neutrino Fluxes from Neutralino annihilations in the Earth and in the Sun}
%============================================================================

The observation of energetic neutrinos from annihilation of neutralinos in
the Sun~\cite{Silk:1985ax,Krauss:1985ks} and/or the 
Earth~\cite{Krauss:1985aa,Freese:1985qw}, is a promising method for indirect detection of
neutralino dark matter (see {\em e.g.}~\cite{Bertone:2004pz,Jungman:1995df}
for a review).

Neutralinos making up the dark matter in the halo of the Galaxy have a small
but finite probability of elastically scattering from a nucleus in 
a given body (the Sun or the Earth). In doing so, neutralinos might
be left with a velocity
smaller than the escape velocity and, thus, become gravitationally bound to the
body. The captured neutralinos settle to the core of the body, via
additional scatterings from nuclei in the body, and eventually annihilate
with one another. 

The pair annihilation of the accumulated neutralinos generates, via decay
of the particles produced in the various annihilation final states, high
energy neutrinos with a differential flux given by:
\beq
\frac{d N_\nu}{d E_\nu}= \frac{ \Gamma_{ann}}{4 \pi d^2} \sum_f{BR_f 
\frac{d N_f}{d E}}
\label{neutrinoflux}
\eeq

Here $d$ is the distance of the detector from the Sun or the center of the Earth, 
$\Gamma_{ann}$ is the annihilation rate of the neutralinos, 
$BR_f$ is their branching ratio
into the final state $\chi \chi \rightarrow f$ and
$d N_f/d E$ is the neutrino spectrum from the decay
of the particles in the final state $f$.

Since neutralinos inside the Sun or the Earth are highly non-relativistic,
their annihilations occur almost at rest. The branching ratios of the
different annihilation channels are discussed in App.~\ref{appannihilation}.

A light neutralino can only annihilate into the light quarks and lepton pairs, which, after decay, give rise to a fairly
soft neutralino spectrum. A more massive $\chi$ can lead to $W^+W^-$, $ZZ$ and
heavier quark pairs, which typically produce a harder differential neutralino flux.
Apart from these {\it fundamental} channels, neutralino annihilations can
produce Higgs bosons or mixed Higgs/gauge boson final states. The Higgs bosons
will, in turn, decay to other Higgses or to one of the ``fundamental'' 
channels~\cite{Ritz:1987mh}.
In our calculations, we have taken into account the fact that the
number of final states containing Higgs bosons is increased in the
\nmssm\ due to the extra CP-even and CP-odd states, $h_3$ and $a_2$,
compared to the \mssm. 

%Let us discuss the other quantities entering the computation in turn.

\subsubsection{Annihilation rate in the Sun and in the Earth}

Neutralinos accumulate in the Sun or the Earth by capture from the halo of
the Galaxy, and are depleted by annihilation and by evaporation.
The evolution equation for the number of neutralinos, $N$, in the Sun or 
the Earth is given by:
\beq
\frac{dN}{dt}=C-C_A N^2 -C_E N
\label{numlsp}
\eeq
where $C$ is the rate of accretion onto the body, the second term is twice the 
annihilation rate and the last term accounts for neutralino evaporation. 

Evaporation has been shown to be important only for neutralinos lighter than 
$3-5\:\gev$~\cite{Griest:1986yu,Gould:1987ju}. The lightest neutralino that we
consider in this paper is on the upper range, $m_\chi \sim 5\:\gev$, so we can
safely neglect the last term in Eq.~(\ref{numlsp}).

We then solve Eq.~(\ref{numlsp}) for $N$, and obtain the annihilation
rate at any given time:
\beq
\Gamma_{ann}=\frac{C}{2} {\rm tanh}^2 \left(t/\tau_A \right)
\label{annsunea}
\eeq
\noi where $\tau_A=1/\sqrt{C C_A}$ is the 
time scale for capture and annihilation equilibrium to occur. We will be
interested in the value of $\Gamma_{ann}$ today, for
$t=t_\odot\simeq1.5\times 10^{17}\: {\rm s}$.

The annihilation rate per effective volume, $C_A$ is given by:
\beq
C_A=\left<\sigma_A v\right> \frac{V_2}{V_1^2},
\label{cav}
\eeq
\noi and $V_j=\left[3 m_{Pl}^2 T/(2 j m_\chi \rho) \right]^{3/2}$
are the effective volumes for the Sun 
($V_j \sim 6.6 \times 10^{28} (j m_{\chi,10})^{-3/2}\: {\rm cm}^3$) or the 
Earth ($V_j \sim 2.3 \times 10^{25} (j m_{\chi,10})^{-3/2}\: {\rm cm}^3$).

The total annihilation rate, $\left<\sigma_A v\right>$, is calculated with
all the contributions at tree level, with the inclusion of the of the
two gluon channel discussed in App.~\ref{appannihilation}.

The accretion rate in the Sun was first calculated in~\cite{Krauss:1985ks} and ~\cite{Press:1985ug},
and for the Earth in~\cite{Freese:1985qw,Krauss:1985aa}. More detailed
evaluations can be found in~\cite{Gould:1987ir}. The results depend on the
velocity dispersion in the halo, the velocity of the Sun with respect to the
halo, the local density of dark matter and the composition of the Sun or
the Earth. For the Sun, we use the 
analytic approximations to the results of~\cite{Gould:1987ir}
that can be found in~\cite{Jungman:1995df}, and the solar model we use is that 
of~\cite{Bahcall:2000nu}, with additional 
abundances taken from~\cite{Grevesse:1998bj}. For the Earth, we 
follow~\cite{Gould:1987ir}.

The capture rate of neutralinos inside the Earth receives an additional contribution from a
subpopulation of neutralinos that scatter on a nucleus located near the
surface of the Sun, and lose enough energy to stay in 
Earth-crossing orbits which, due to planetary perturbations, do not intersect with the Sun~\cite{Damour:1998rh}.
This addition to the local density of dark matter in the Earth has a
characteristic velocity that more closely matches the escape velocity from the
Earth than the background halo population, enhancing the resonant capture
off elements such as iron. This effect is more important for the light,
$m_\chi \lesssim 100\: \gev$, neutralinos that we are considering and we thus
take it into account when computing capture rates in the Earth.

The capture of neutralinos in the Sun or the Earth depends on the elastic
scattering cross sections with the nuclei that make up the body. These
cross sections can be derived~\cite{Jungman:1995df}
from the nucleon (proton or neutron) cross
sections that are discussed, for the \nmssm, in App.~\ref{appelastic}. We
have included both spin-independent and spin-dependent terms in our
computations, the latter being potentially important to evaluate
the accretion in the Sun.

\begin{figure}
\begin{center}
\begin{tabular}{c@{\hspace{1.5cm}}c}
  \includegraphics[width=0.45\textwidth]{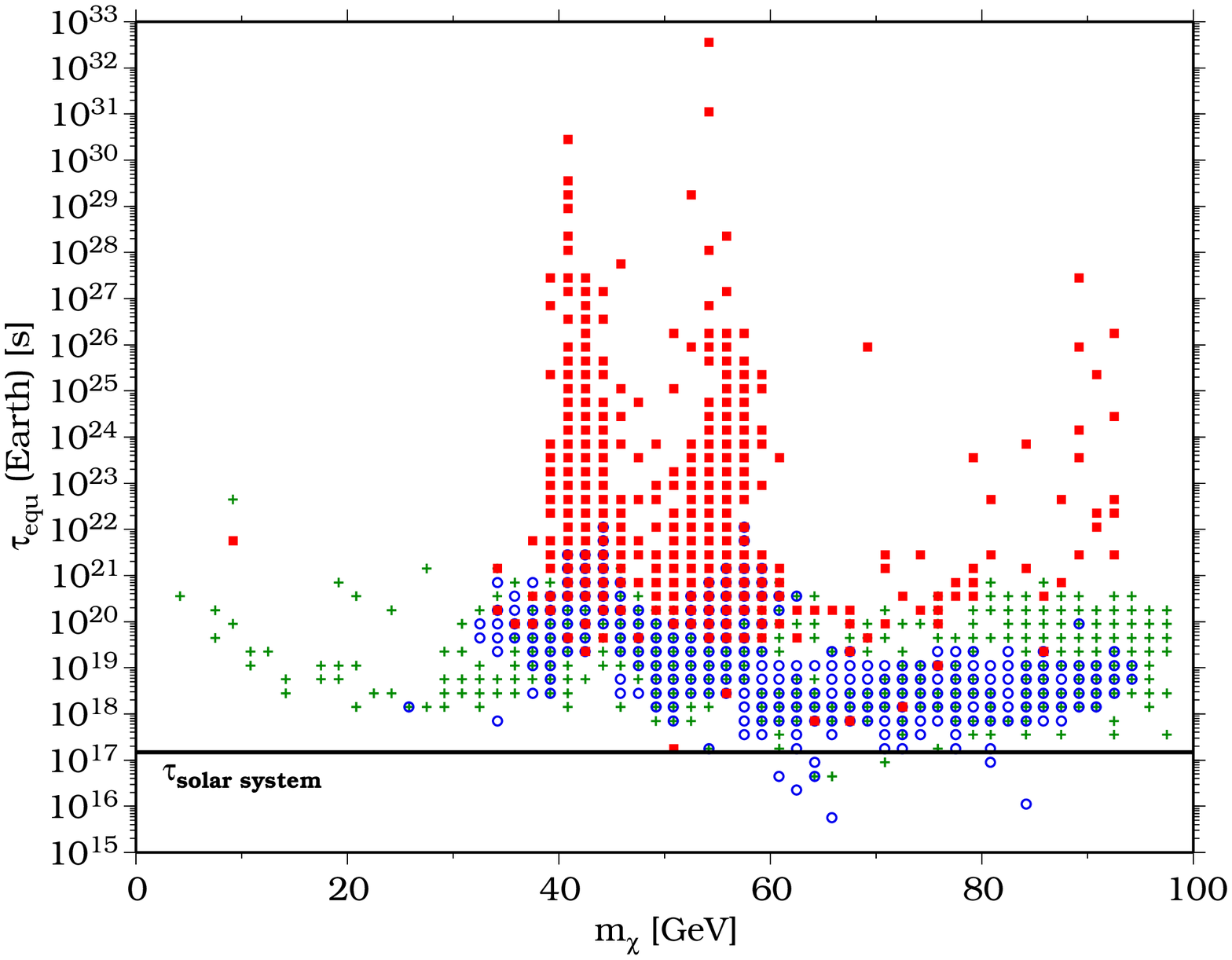}&
  \includegraphics[width=0.45\textwidth]{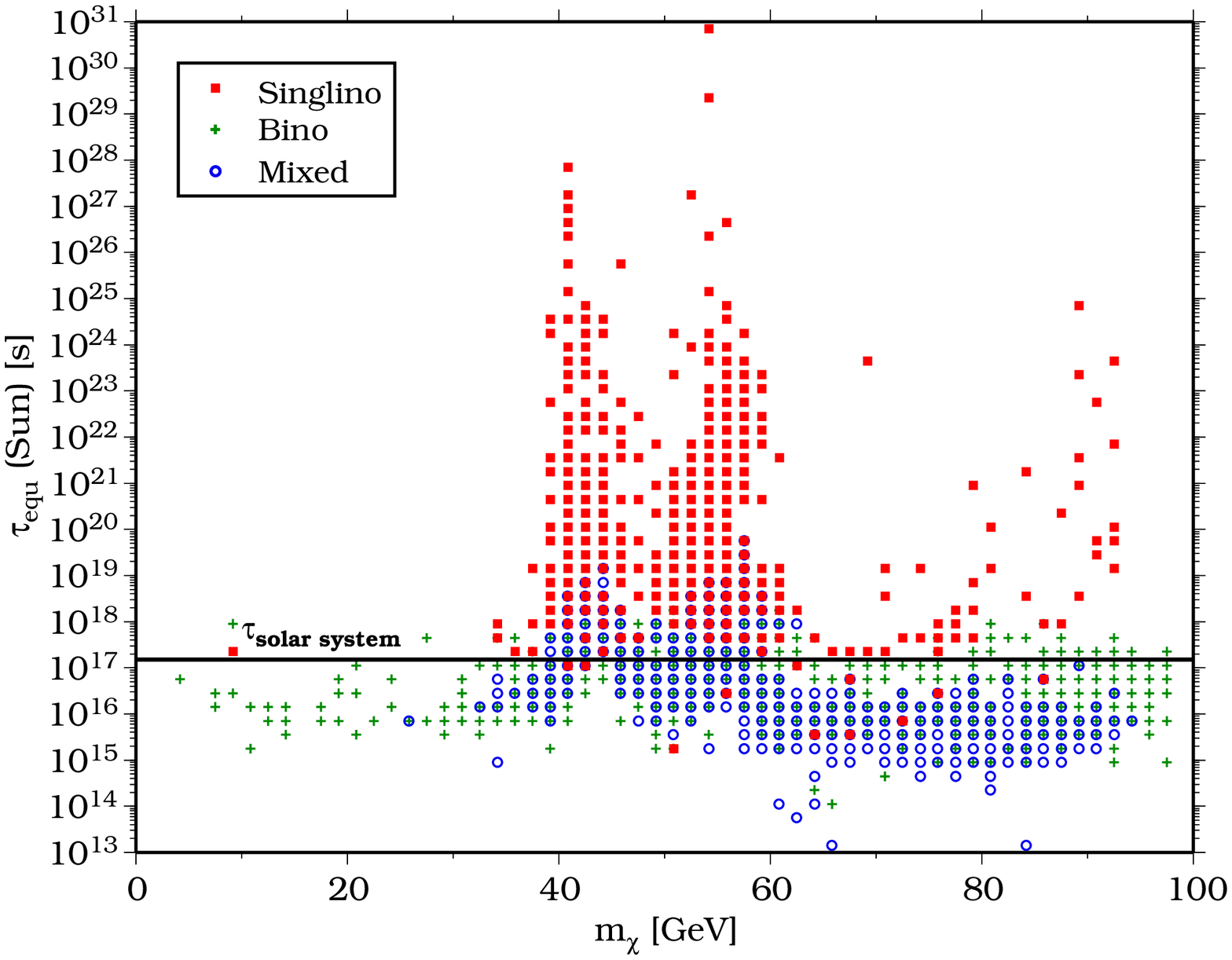}
\end{tabular}
\end{center}
\caption{Equilibrium times between capture and annihilation in
the Earth (left) and the Sun (right). Models below the line marking
the age of the solar system have attained equilibrium and will have
an $\Gamma_{ann}\sim C/2$}
\label{tausunearth}
\end{figure}

We show in Fig.~\ref{tausunearth} the equilibrium time between capture and
annihilation in the Earth and in the Sun. Most bino and mixed-like
neutralinos have reached equilibrium and 
${\rm tanh}^2 \left(t_\odot/\tau_A \right) \sim 1$ in Eq.~(\ref{annsunea}). In the
Earth, featuring a shallower gravitational potential well, 
equilibrium has only been reached by a few mixed-like
neutralinos, and the annihilation rate will be below $C/2$. The emission
region is however much closer to the detector, and, as we will see 
below, and in contrast to the usual situation in the MSSM, the constraints from the Earth are more stringent than those from
the Sun.

\subsubsection{Muon fluxes}

The neutralino annihilation products will hadronize and/or decay giving rise
to high energy neutrinos, $E_\nu \lesssim m_\chi$, which may be
detected in a neutrino telescope by measuring the upward-going muons produced
by $\nu_\mu$ and $\bar{\nu}_\mu$ interactions in the rock below the detector.

The precise
determination of the secondary neutrino spectrum is a difficult problem that
calls for dedicated Monte Carlo simulations of the hadronization and energy
losses in the medium of the body. We have used the results
of~\cite{Bergstrom:1996kp} and adapted the relevant routines in 
\ds~\cite{Gondolo:2004sc}, to 
take into account the additional Higgses present in the \nmssm. Spectra are given for six {\it fundamental} channels, 
$\chi \chi \rightarrow c \bar{c}, \: b \bar{b}, \: t \bar{t}, \: \tau \bar{\tau}, \: W^+W^-, \: ZZ$, which are also used for the Higgs and Higgs/gauge
boson final states by following the decay chain until one of the fundamental
channels is reached.

The muon yields in~\cite{Bergstrom:1996kp} include the effects of 
hadronization/decay of the annihilation products, $\nu$ interactions on their
way out of the Sun and near the detector, and of the multiple Coulomb
scattering of the $\mu$ on its way to the detector. A similar study was
done in~\cite{Bottino:1994xp}. 

The effects of oscillations in the propagation of the neutrinos through the
Sun have been discussed in~\cite{Roulet:1995qb}.
More recently, the full spectra of all neutrino flavors including additional
channels, such as light quarks and gluons, and accounting for oscillations
and $\nu_\tau$-regeneration were given in~\cite{Cirelli:2005gh}. The combined
effect amounts to a ${\cal O} (0.1-10)$ correction which is comparable to
astrophysical uncertainties. We do not include these effects here, although
if an anomalous $\nu$ signal were discovered, it would be then interesting to 
try to reconstruct
the mass and branching ratios of the would-be neutralinos~\cite{Cirelli:2005gh}.

\begin{figure}
\begin{center}
\begin{tabular}{c@{\hspace{1.5cm}}c}
  \includegraphics[width=0.45\textwidth]{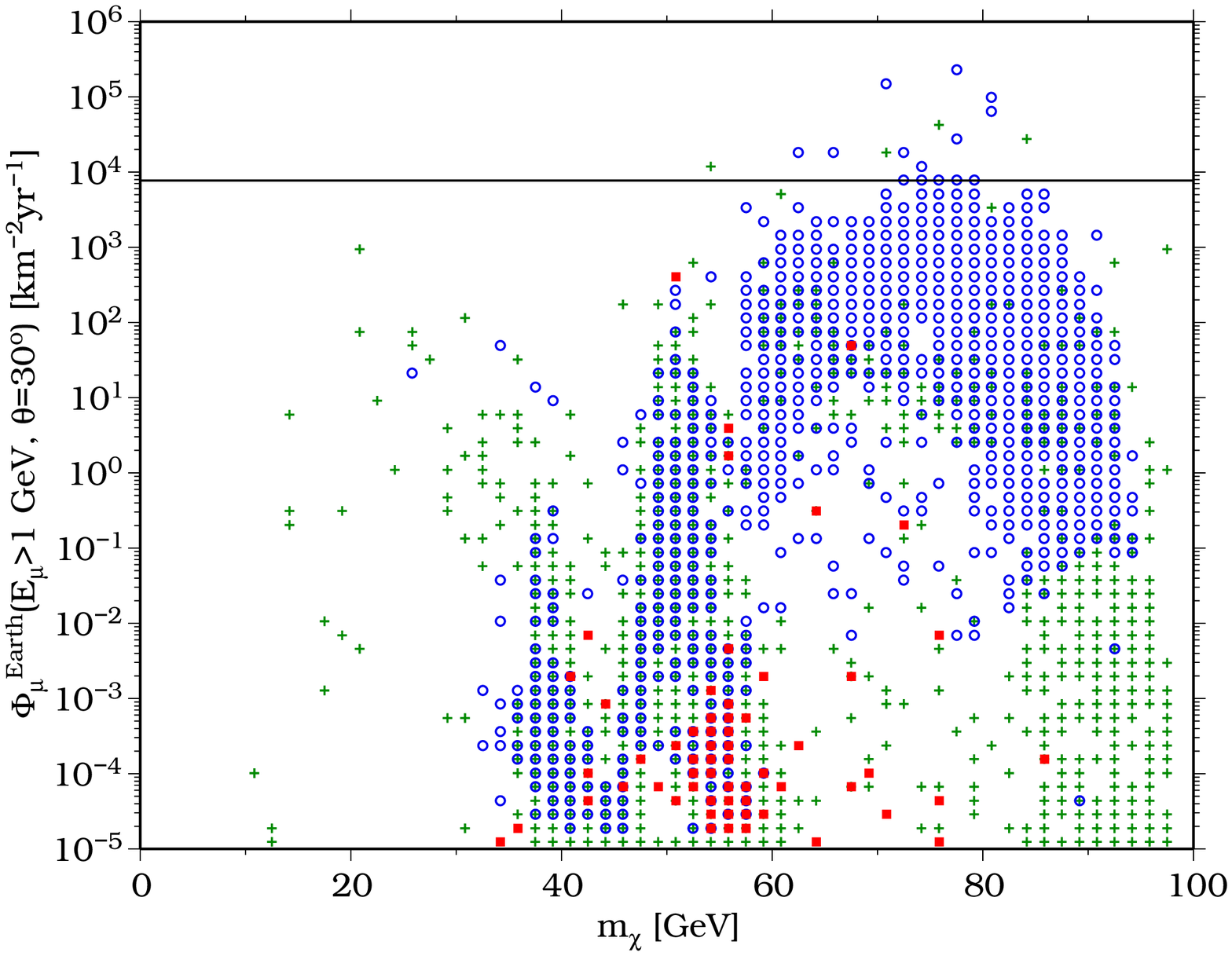}&
  \includegraphics[width=0.45\textwidth]{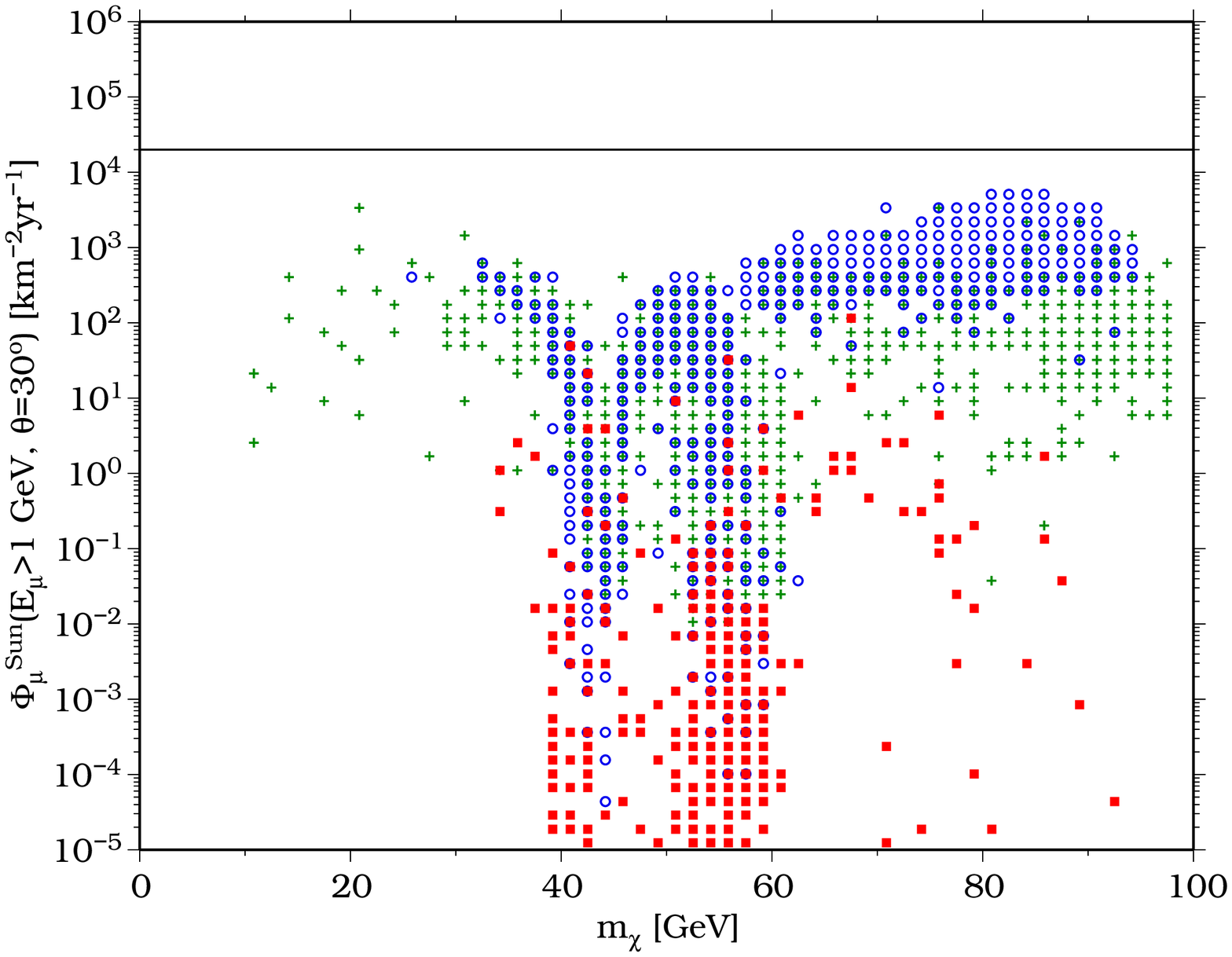}
\end{tabular}
\end{center}
\caption{Integrated muon fluxes above $E_\mu \ge 1 \gev$ from the 
Earth (left) and the Sun (right). The horizontal line displays the
MACRO bound~\cite{Ambrosio:1998qj}.}
\label{muonflux}
\end{figure}

Fig.~\ref{muonflux} shows the muon fluxes for $\chi \chi$ annihilation
from the Earth and from the Sun. We show integrated fluxes above a
threshold energy of $1\:\gev$ and the horizontal line represents the MACRO
limit~\cite{Ambrosio:1998qj} which is comparable to that of
Super-Kamiokande~\cite{Desai:2004pq}. To be able to constrain the low mass
neutralinos considered in this work, it is crucial for the detector to have
a low threshold. Of all the forthcoming facilities, 
ANTARES~\cite{Aguilar:2003us} seems to be the most
promising one, with an advertised threshold of $E_\mu \sim 10\: \gev$ and
a target sensitivity of $100-1000\: {\rm km}^{-2} {\rm yr}^{-1}$, 
it should be able
to detect or further constrain those models with $m_\chi \gtrsim 20\: \gev$.
Larger facilities like ICECUBE~\cite{Achterberg:2006md}
have sparser instrumentation, which
increases the threshold to $E_\mu \gtrsim 100\: \gev$, above the mass
range in which we are interested.

In Fig.~\ref{muonflux}, we considered a half-aperture of $\theta=30^\circ$.
A cone of this size should contain most of the muons from annihilations of
even the lightest neutralinos. For moderately larger masses, a
smaller aperture could improve the signal to noise ratio by reducing the
background while still collecting most of the signal, and the limits 
could be improved by an optimized analysis.

It is encouraging, however, that present muon fluxes due to capture in the Earth, presumably in part due to the enhancement in the density of light neutralinos in solar system orbits, are currently able to
rule out a few models with moderate masses, $m_\chi \sim 60-80\: \gev$, and that an
order of magnitude improvement in sensitivity as expected with the ANTARES telescope, should
enable to access a sizable part of the parameter space by looking at signals
from both the Sun and the Earth. On the other hand, singlino-like neutralinos
show suppressed muon fluxes and prospects for their detection seem more
remote.

\subsection{Solar Physics Bounds}
\label{solarbounds}

Energy transport by neutralinos could have relevant effects on the Sun,
producing an isothermal core and reducing the Sun central temperature, $T_c$.
WIMPs with masses of a few $\gev$ and elastic scattering
cross-sections around $\sigma_{el}\sim 10^{-36}\:{\rm cm}^2$, were considered
some time ago as being able to reduce the solar neutrino flux, hence
solving the solar neutrino problem ~\cite{Krauss:1984a,Spergel:1984re,
Krauss:1985ks,Faulkner:1985rm}. It has, since then, been realized that the
solar neutrino problem cannot be solved by simply reducing $T_c$, and this hypothesis was abandoned.

On the other hand, our knowledge of the solar interior has advanced to a
point where stellar evolution theory in combination with observational
data could provide information on the existence and properties of the particles
constituting the dark matter. The sound speed in the Sun is known with an
accuracy of roughly $0.1 \%$ through helioseismic 
data~\cite{Degl'Innocenti:1996ev}, and the measurement of the
neutrino flux from ${}^8 B$ decay has enabled the determination of $T_c$
at the percent level~\cite{Fiorentini:2001et}.

The variations in the sound speed induced by dark matter particles were
considered in~\cite{Lopes:2001ra} and, together with the influence on the 
boron neutrino flux~\cite{Lopes:2001ig},
were claimed to exclude WIMPs below $m_\chi \lesssim 60\: \gev$. This stringent
conclusions were due, according to~\cite{Bottino:2002pd}, to an unrealistic
extrapolation of the helioseismic data down to the central regions of the
Sun. Neutralinos as light as $m_\chi \sim 30\: \gev$ were shown to be in
accord with helioseismology and also to leave the neutrino fluxes unchanged,
since the central temperature was only being modified in a small region
around the center of the Sun. 

It is nonetheless of interest to consider
the influence on the solar energy transport of neutralinos within the \nmssm.
Apart from changes in the capture rates, the mass of the neutralinos we are considering here dwell well below $m_\chi \sim 30\: \gev$, creating
a larger isothermal core with potential observable effects.

Energy transport in the Sun can occur by diffusion or in a non-local manner.
The prevalence of either regime is determined by the {\it Knudsen} number,
which is the ratio of the mean free path of the weakly interacting neutralino
in the multicomponent baryonic background to the scale length of the system:
\beq
Kn \equiv \left(L \sum_i{\sigma_i n_i} \right)^{-1},
\label{knudsen}
\eeq
where the sum runs over the chemical elements in the Sun.

For neutralinos in the Sun, the relevant geometric dimension is the scale
height of the neutralino cloud in the central region, which can be
approximated by:
\beq
L = r_\chi \sim 0.13 R_\odot \sqrt{\frac{1\:\gev}{m_\chi}}.
\label{rchi}
\eeq

When the mean free path is short compared to $r_\chi$, energy is transported
by thermal conduction and the relevant Boltzmann collision equation has been carefully
studied in~\cite{Gould:1989hm}. We will be mostly interested in the
opposite regime, the Knudsen limit, when $Kn \gg 1$ and the particles orbit
many times in the Sun between interactions with nuclei. An analytic
approximation for this case was presented in~\cite{Spergel:1984re},
although Monte Carlo simulations~\cite{Nauenberg:1986em} revealed that it
overestimated the neutralino luminosity by a factor of a few. This
was subsequently confirmed, and the source of the discrepancy attributed
to the deviation from isotropy of the neutralino 
distribution~\cite{Gould:1989ez}. With this caveat in mind, it will suffice, for
our purposes, to estimate the neutralino luminosity using the results
of~\cite{Spergel:1984re}.

We asserted above that the neutralinos will be transferring energy in the
Knudsen regime: let us show now that this is indeed the case. The critical cross
section for an interaction to occur in a solar radius can be estimated
as:
\beq
\sigma_c \sim \frac{m_p}{M_\odot} R_\odot^2 \sim 4 \times 10^{-36} {\rm cm}^2
\label{sigmacrit}
\eeq

\begin{figure}[hbt]
\begin{center}
\includegraphics[width=0.5\textwidth]{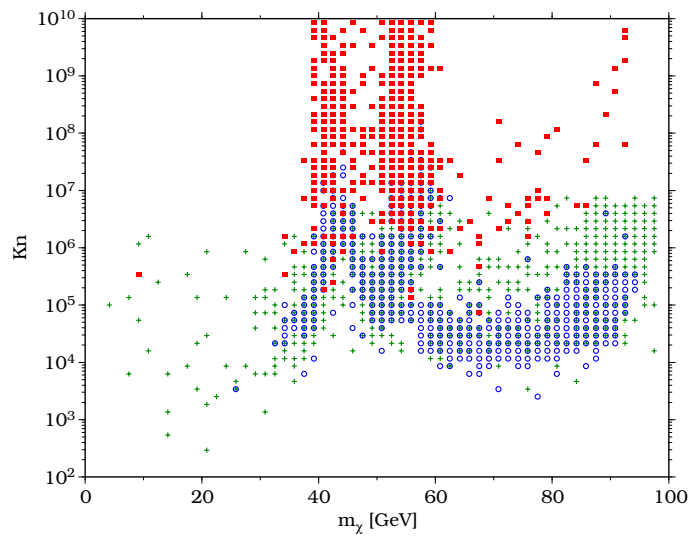}
\end{center}
\caption{Knudsen number, Eq.~(\ref{knudsen}), at a distance $r_\chi$}.
\label{knud}
\end{figure}

The $\chi-n,p$ elastic scattering cross sections 
that we obtain using the results in
App.~\ref{appelastic} fall a few orders of magnitude below $\sigma_c$. For
some mixed-like neutralinos we get values as large as 
$2 \times 10^{-39}\ {\rm cm}^2$,
an order of magnitude larger than for bino-like neutralinos and some
three orders of magnitude above those of singlinos. Hence, the neutralinos
will travel over distances larger than $10^3 R_\odot$, corresponding
to Knudsen parameters in the range $Kn \gtrsim 10^3$. We show the parameter
in Fig.~\ref{knud} at a distance $r_\chi$ from the center of the Sun,
which is where the neutralino luminosity is expected to 
peak~\cite{Gould:1989ez}\footnote{In this respect, the quantity
$\delta$ used in~\cite{Bottino:2002pd} does not seem appropriate to characterize
neutralino energy transport, since it involves its luminosity at the
center of the Sun.}.

The energy transport is most effective in the region 
$Kn \sim 0.4$~\cite{Gould:1989hm,Gould:1989ez}, far below the values
depicted in Fig.~\ref{knud}, so we expect the neutralino luminosity
to be a fraction of the total solar luminosity ${\cal L}_\odot$.
Indeed, we can obtain a rough estimate of ${\cal L}_\chi$, by adapting
Eqs.~(2.8-2.10) in~\cite{Spergel:1984re} to account for the different species
of nuclei in the Sun, and assuming the neutralino luminosity is confined to a region of size $r_{\chi}$:
\beq
{\cal L}_\chi \sim \frac{N_\chi R_\odot^2 \sigma_c}{r_\chi^4 Kn} \sqrt{\frac{1 \: \gev}
{m_\chi}} 4.1 \times 10^{12} {\cal L}_\odot,
\label{lchi}
\eeq
where the number of neutralinos in the Sun is given by
$N_\chi = C \tau_A {\rm tanh} (t_\odot/\tau_A)$.

\begin{figure}[hbt]
\begin{center}
\includegraphics[width=0.5\textwidth]{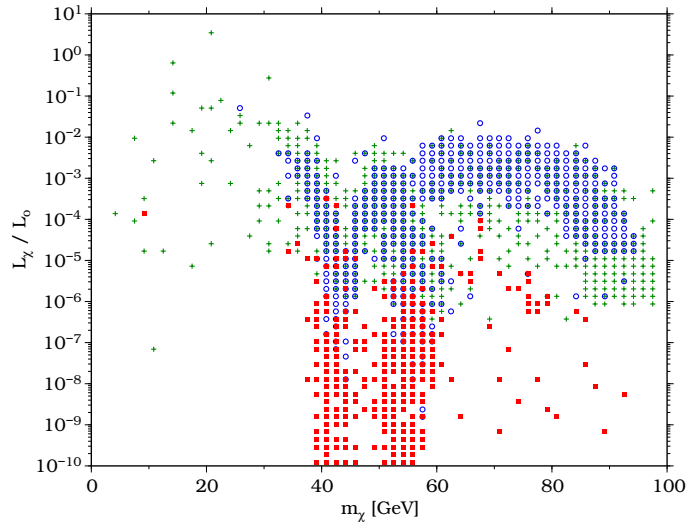}
\end{center}
\caption{Neutralino luminosity in units of ${\cal L}_\odot$.}
\label{lumchi}
\end{figure}

Looking at Fig.~\ref{lumchi}, where we see that most models contribute a tiny
fraction of the solar luminosity. However for the lightest bino-like
neutralinos, $m_\chi \lesssim 30\ \gev$, the neutralino luminosity may be comparable
to the total solar luminosity, and may thus already be disfavored.

We have already mentioned that the neutralino luminosity might be overestimated
by a factor of ${\cal O}(10)$. On the other hand, the neutralino luminosity
is not directly observable, and it is not unconceivable that neutralinos
giving a smallish fraction of the total luminosity, but having a small mass
and, hence, a sizable isothermal core, might modify the boron neutrino
fluxes appreciably. As pointed out in~\cite{Spergel:1984re}, a 
$20\ \gev$ neutralino
carrying only $10^{-2} {\cal L}_\odot$, could be responsible for the
transfer of up to $50\%$ of the energy in the inner region bounded by
$r_\chi$. Computing the actual modifications in neutrino fluxes
and/or helioseismic data, would require the generation of self-consistent
solar models with the neutralino transport taken into account. Although
this is beyond the scope of the present work, the present estimates suggest
that this task may deserve further investigation for light NMSSM neutralinos.

\subsection{Antimatter from Neutralino annihilations in the Galactic Halo}

%==========================================================================

Neutralino pair-annihilations in the galactic halo can produce, through the hadronization or decay of the underlying elementary constituents arising from the annihilation process, antimatter in the form of positrons and of hadronic stable antimatter states like antiprotons and antideuterons. The abundance of antimatter fluxes produced in neutralino pair-annihilations not only depends upon the particle physics nature of neutralinos, but also on various astrophysical factors. The latter--including the structure of the dark matter galactic halo, the propagation of cosmic rays in the Galaxy, the effects of solar modulation--induce some amount of uncertainty in the flux computation. Further, while in the case of low-energy antideuterons the cosmic ray background can be suppressed at a level where the detection of even a single antideuteron can be a signal for new physics and potentially for dark matter annihilations in the halo, for positrons and antiprotons the background is large.  While this latter background is, to some extent, understood, it has to be properly incorporated and estimated if one is to be able to extract a possible dark matter annihilation signal from the data. 

As far as the dark matter distribution in the galactic halo is concerned, we resort here to the strategy outlined 
in Ref.~\cite{Profumo:2004ty} (the reader is referred to 
Ref.~\cite{pierohalos,Edsjo:2004pf,Profumo:2004at} for more details). We consider two extreme possibilities for the structure of the dark matter halo.
In the first scenario, the central cusp in the dark matter halo, 
as seen in numerical simulations, is smoothed out by a significant 
heating of the cold particles~\cite{elzant}, leading to a cored 
density distribution, which has been modeled by the so called 
{\em Burkert profile}~\cite{burkert},
\begin{equation}\label{eq:burkert}
\rho_B(r)=\frac{\rho_B^0}{(1+r/a)\left(1+(r/a)^2\right)}.
\end{equation}
Here, the length scale parameter 
has been set to $a=11.7$ kpc, while the normalization
$\rho_B^0$ is adjusted to reproduce the local halo density at the 
Earth position to $\rho_B(r_0)=0.34 \ {\rm GeV\ cm}^{-3}$~\cite{pierohalos}. 
We refer to this model as to the {\em Burkert Halo Model}.
It has been successfully tested against a large sample of rotation 
curves of spiral galaxies~\cite{salucci}. 
%always giving satisfactorily accurate 
%fits  

In the second scenario we consider here, 
baryon infall causes a progressive 
deepening of the gravitational potential well at the center of the galaxy, 
resulting in an increasingly higher concentration of dark matter particles. 
In the circular orbit approximation~\cite{blumental,ulliobh}, 
this adiabatic contraction limit has been worked out starting from the 
N03 profile proposed in Ref.~\cite{n03}; the resulting spherical profile, 
which has no closed analytical form, roughly follows, in the inner 
galactic regions, the behavior of the profile proposed by 
Moore {\it et al.},\cite{moore}, approximately scaling as $r^{-1.5}$ in the 
innermost regions, and features a local dark matter density 
$\rho_{N03}(r_0)=0.38 \ {\rm GeV\ cm}^{-3}$. 
We dub this setup as the {\em Adiabatically Contracted N03 Halo Model}.

The parameters for both models have been chosen to reproduce a variety 
%implementing a set 
of dynamical information, ranging from the constraints 
stemming from the motion of stars in the sun's neighborhood, 
total mass estimates from the motion of the outer satellites, 
and consistency with the Milky Way rotation curve and measures of the 
optical depth toward the galactic bulge~\cite{pierohalos,Edsjo:2004pf}.
Both models have been included 
in the latest public release of the \ds 
package~\cite{Gondolo:2004sc}.

The antimatter yields from neutralino annihilation are then computed following the procedure 
outlined in Ref.~\cite{Profumo:2004ty}. 
We calculate the neutralino annihilation rates to $\bar{p}$ and
$\bar{n}$ 
using the 
{\sc Pythia} 6.154 Monte Carlo code~\cite{pythia} as implemented in 
\ds~\cite{Gondolo:2004sc}, and then deduce the $\overline{D}$ 
yield using the prescription suggested in Ref.~\cite{dbar}.
% to convert from the $\bar p$ and $\bar n$ yields. 
The propagation of charged 
cosmic rays through the galactic magnetic fields is worked out through 
an effective two-dimensional diffusion model in the steady state 
approximation \cite{Bergstrom:1999jc}, while solar modulation effects were implemented through the 
analytical force-field approximation of 
Gleeson and Axford~\cite{GleesonAxford}.
The solar modulation parameter $\Phi_F$ is computed from the proton 
cosmic-ray fluxes, and assumed to be charge-independent. 
The values of $\Phi_F$ we make use of refer to a 
putative average of the solar activity over the three years of data 
taking of the recently launched Payload for Antimatter Matter Exploration and Light-nuclei Astrophysics (PAMELA) experiment~\cite{pamelanew} for positrons and antiprotons, and over the estimated period of data-taking for the General Anti-Particle Spectrometer (GAPS) experiment in the case of antideuterons.

%------------------------------------------------------------------------------
\begin{figure}[t]
\begin{center}
\includegraphics[width=0.9\textwidth]{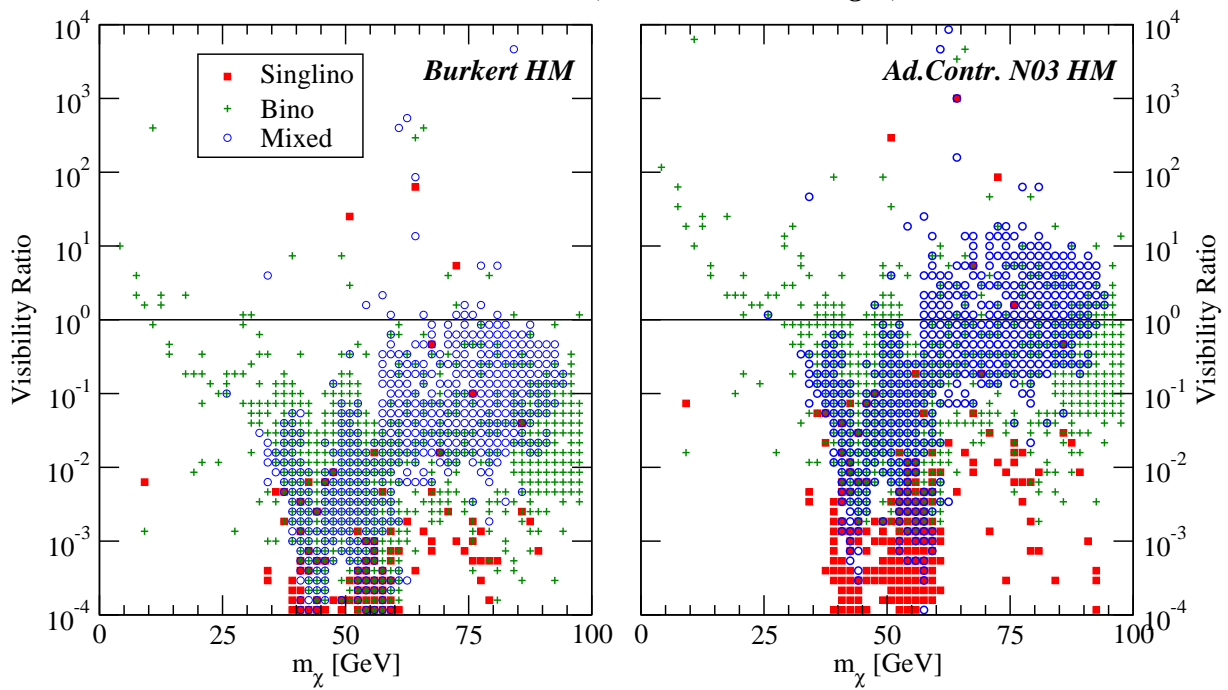}
\end{center}
\caption{The expected number of antideuterons detectable with an ultra-long duration balloon-borne GAPS-type experiment, as a function of the lightest neutralino mass. In the left panel we adopt a Burkert halo model, while in the right panel we make use of an adiabatically contracted N03 halo profile. The conventions for the various neutralino types follow those of Fig.~\ref{fignmssm}.}
\label{fig:dbar}
\end{figure}
%------------------------------------------------------------------------------
For antideuterons
we consider the reach of the proposed {\em general} antiparticle 
spectrometer (GAPS)~\cite{Mori:2001dv,gapsnew} in an ultra-long duration balloon-borne (ULDB) mission, tuned to look for 
antideuterons in the very low kinetic energy interval from 0.1 
to 0.25 GeV per nucleon. As described in Ref.~\cite{dbarprofumo}, in fact, this experimental setting would allow one to safely neglect the background from secondary and tertiary cosmic-ray-produced antideuterons, unlike a satellite-borne mission: the detection of a single low-energy antideuteron would then be a clean signature of an exotic antideuteron source (including, but not limited to, galactic dark matter annihilation).
We set the value of the solar modulation 
parameter $\Phi_F$ at the value corresponding to the projected year for the balloon-borne GAPS mission, around 2011.  
The resulting sensitivity of 
GAPS has been determined to be of the level of 
$3\times 10^{-8} {\rm m}^{-2}{\rm sr}^{-1} {\rm GeV}^{-1} {\rm s}^{-1}$
~\cite{gapsnew,dbarprofumo}. 

%------------------------------------------------------------------------------
\begin{figure}[t]
\begin{center}
\includegraphics[width=0.9\textwidth]{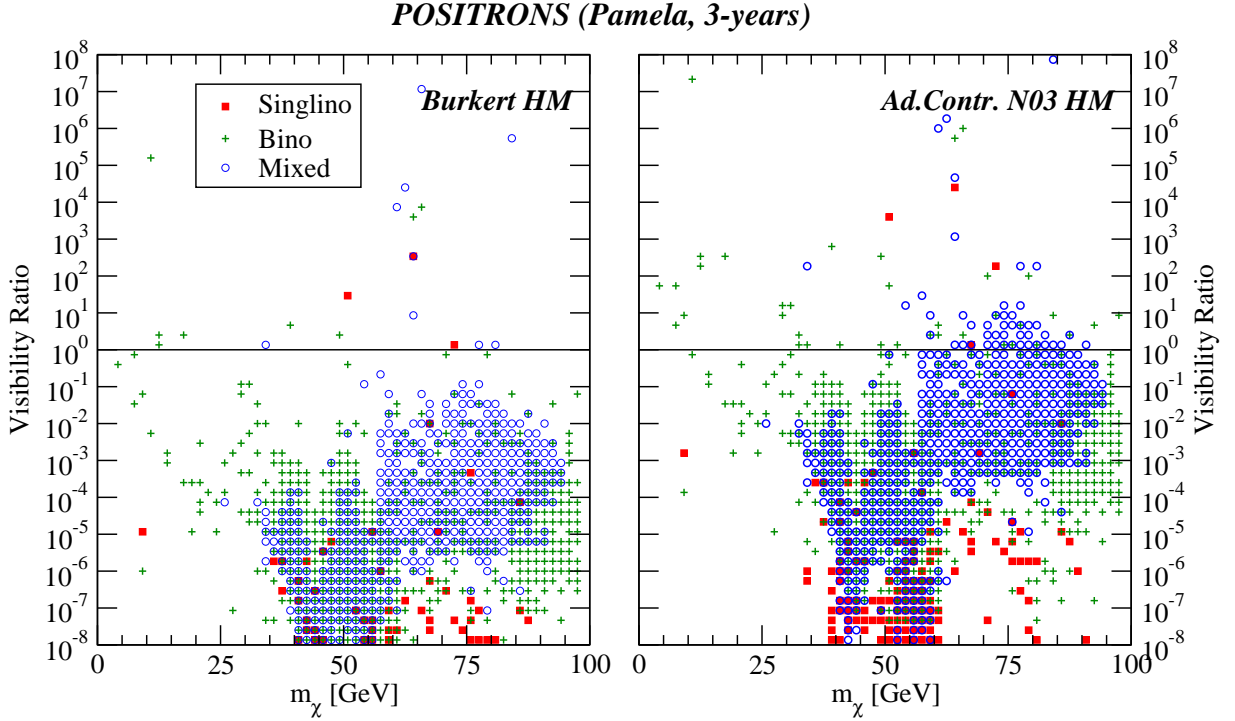}
\end{center}
\caption{The Visibility Ratio for positrons, as defined in Eq.~(\protect{\ref{eq:vr}}), as a function of the lightest neutralino mass. In the left panel we adopt a Burkert halo model, while in the right panel we make use of an adiabatically contracted N03 halo profile. The conventions for the various neutralino types are as in Fig.~\ref{fignmssm}.}
\label{fig:eplus}
\end{figure}
%------------------------------------------------------------------------------
To evaluate the sensitivity of the PAMELA antimatter search experiment, we adopt the statistical treatment of the
antimatter yields introduced in Ref.~\cite{Profumo:2004ty} (an analogous
approach has been proposed for cosmic positron searches
~\cite{Hooper:2004bq}). Motivated by the fact that the signal is much
smaller than the background, we introduce a quantity which weighs the signal's
``statistical significance, summed over the energy bins'',
\begin{equation}\label{eq:iphi}
I_\phi=\int_{T_{\rm min}}^{T_{\rm max}}\frac{\left(\phi_s(E)\right)^2}
{\phi_b(E)}{\rm d}E,
\end{equation}
where $\phi_s(E)$ and $\phi_b(E)$ respectively represent the antimatter 
differential fluxes 
from neutralino annihilations and from the background at antiparticles' 
kinetic energy $E$, and $T_{\rm min,\ max}$ 
correspond to the antiparticle's maximal and minimal kinetic energies to which 
a given experiment is sensitive (in the case of the PAMELA 
experiment~\cite{pamelanew}, 
$T^{e^+}_{\rm min}=50$ MeV, $T^{e^+}_{\rm max}=270$ GeV, 
$T^{\bar p}_{\rm min}=80$
MeV and $T^{\bar p}_{\rm max}=190$ GeV). It can be easily verified that Eq.~(\ref{eq:iphi}) reproduces, in the large-number-of-bins limit, the excess $\chi^2$ from an exotic contribution in the fit to the prospect antimatter fluxes.
We compute the primary component, $\phi_s$, with the \ds package, 
interfaced with a subroutine implementing the diffusion and solar modulation 
models outlined above. The background flux $\phi_b$ has been calculated with 
the {\sc Galprop} package~\cite{galprop}, with the same propagation and solar modulation parameter choices 
employed to compute the signal. 

Given an experimental facility with a geometrical factor
(acceptance) $A$ and a total data-taking time $T$, it has been 
shown~\cite{Profumo:2004ty} that, in the limit of a large number of energy 
bins and of high precision secondary ({\em i.e.} background) flux 
determination, a SUSY model giving a primary antimatter flux $\phi_s$ 
can be discriminated at the 95\% C.L. if
\begin{equation}
I_\phi(\phi_s)\cdot A\cdot T>(\chi^2)^{95\%}_{\rm n_b},
\end{equation}
where $(\chi^2)^{95\%}_{\rm n_b}$ stands for the 95\% C.L. $\chi^2$ with 
${\rm n_b}$ degrees of freedom. 
For the PAMELA experiment, where $A=24.5\ {\rm cm}^2\ 
{\rm sr}$, $T$=3 years and ${\rm n_b}\simeq60$ we get the following 
discrimination condition~\cite{Profumo:2004ty}
\begin{equation}
I_\phi(\phi_s)>\frac{(\chi^2)^{95\%}_{\rm n_b}}{A\cdot T}
\equiv I_\phi^{\rm 3y,\ PAMELA,\ 95\%}\simeq3.2\times 10^{-8} 
\ {\rm cm}^{-2} {\rm sr}^{-1} {\rm s}^{-1}
\end{equation}
which is approximately valid for both positrons and antiprotons 
(though in the latter case the PAMELA experiment 
is expected to do slightly better). 
As a rule of thumb, the analogous quantity for AMS-02 should improve at 
least by one order of magnitude~\cite{Feng:2000zu}. 
In our plots, we will show, for both antiprotons and positrons, the following ``Visibility Ratio''
\begin{equation}\label{eq:vr}
{\rm Visibility\ Ratio}\equiv I_\phi^{\bar p,e^+}/
I_\phi^{\rm 3y,\ PAMELA,\ 95\%}.
\end{equation}

%==========================================================================

\begin{figure}[t]
\begin{center}
\includegraphics[width=0.9\textwidth]{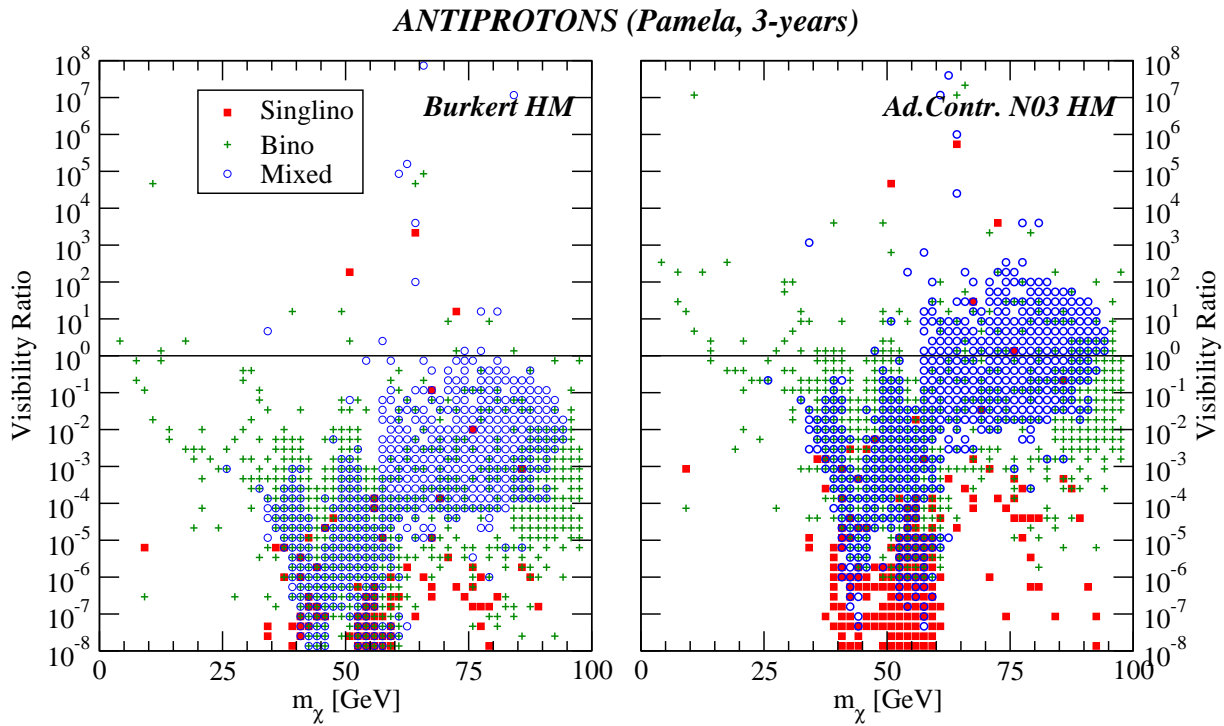}
\end{center}
\caption{The Visibility Ratio for antiprotons, as defined in Eq.~(\protect{\ref{eq:vr}}), as a function of the lightest neutralino mass. In the left panel we adopt a Burkert halo model, while in the right panel we make use of an adiabatically contracted N03 halo profile. The conventions for the various neutralino types are as in Fig.~\ref{fignmssm}.}
\label{fig:pbar}
\end{figure}

We show our results on the prospects for detecting a WIMP pair annihilation signature in the various above mentioned antimatter channels in Fig.~\ref{fig:dbar}-\ref{fig:pbar}. As in the previous figures, we indicate singlino-like models with red squares, bino-like with green pluses and mixed singlino-bino models with empty blue circles. Models lying {\em above} the horizontal lines are expected to give a detectable signature at the experiments discussed above. In all three figures, we adopt the Burkert Halo Model in the left panels and the adiabatically contracted N03 Halo Model in the right panels. As a general comment, switching from the conservative Burkert profile to the more optimistic adiabatically contracted halo profile causes an increase in the fluxes of around one order of magnitude (notice that, in terms of the Visibility Ratio, Eq.~(\ref{eq:vr}), for antiprotons and positrons, which depends on the {\em square} of the signal flux, this translates into a {\em two} orders of magnitude increase). 

%------------------------------------------------------------------------------
%------------------------------------------------------------------------------
We start showing, in Fig.~\ref{fig:dbar}, the Visibility Ratio for antideuterons, effectively given by the expected number of detected antideuterons at an ULDB GAPS mission. As alluded above, this experimental setup is virtually devoid of cosmic ray background, hence the detection of even only one $\bar D$ can be regarded as a ``{\em signal}''. We notice that in general, low mass neutralinos, peculiar of the NMSSM setup under consideration here, yield a sizable flux of low-energy antideuterons. With some exceptions, singlino-like neutralinos produce an insufficient flux of $\bar D$, while the most promising models are mixed singlino-bino models with a mass in the range $55\lesssim m_\chi/{\rm GeV}\lesssim 95$.

Fig.~\ref{fig:eplus} and \ref{fig:pbar} respectively show the Visibility Ratios for positrons and for antiprotons. As a general comment, we point out that in the present setup antiprotons stand as a more promising channel to effectively disentangle an exotic signal. As for the case of antideuterons, low mass models are again expected to give a sizable antimatter yield. While we do find some instances of singlino-like neutralinos that can give large antimatter fluxes, in general we find that the antiproton and positron yield from singlinos is not particularly promising. On the other hand, mixed models, peculiar to the NMSSM, give in general large fluxes, and a significant portion of the models will be tested by the results from the space-based PAMELA experiment on a time scale of three years (or by AMS-02 in a much shorter time scale).

We also computed the constraints from {\em current} antiproton \cite{currpbar} and positron \cite{curreplus} flux measurements, in terms of the $\chi^2$ to the data of the sum of the background and the signal. Using this criterion, we find that models featuring an antiproton Visibility Ratio larger than $\sim10$ are generically conflicting with current data, and so are models giving a positron Visibility Ratio larger than $\sim300$. However, one should keep in mind that the background we use in our computation can be somewhat lowered without conflicting with cosmic ray propagation models; in a more conservative approach, asking that the signal alone does not exceed the measured antiproton flux, rules out only models with Visibility Ratios larger than $\sim100$ ($\sim1000$ in the case of positrons).

\subsection{The Monochromatic Gamma-ray Flux}

Neutralinos can pair annihilate in the Galaxy or in dark matter concentrations outside the Galaxy yielding a coherent and directional flux of gamma rays; two components add up in the total gamma-ray yield expected from neutralino pair-annihilations: a continuum part, extending up to gamma-ray energies $E_\gamma\lesssim m_\chi$, generated by annihilation products radiation and from decays of, {\em e.g.}, $\pi^0\rightarrow\gamma\gamma$, and (possibly more than one) monochromatic lines, in loop-induced direct decays to, {\em e.g.}, $\gamma\gamma$, $Z\gamma$ or $H\gamma$ final states. Among the latter, the brightest, and the one which occurs in any supersymmetric framework (the others being potentially kinematically forbidden) is often that associated to the $\gamma\gamma$ final state. Since the possibility of unambiguously disentangling the continuum gamma-ray contribution from the background is known to be observationally extremely challenging (see {\em e.g.} the recent analyses in Ref.~\cite{gammaraysref1, gammaraysref2, Zaharijas:2006qb}), and in view of our expectations on the size of the $\gamma\gamma$ annihilation channel in the NMSSM, as anticipated in the Introduction, we shall concentrate here on the monochromatic gamma-ray line from radiative annihilation of neutralinos into two photons, at an energy $E_\gamma=m_\chi$.

As well known, the estimate of the gamma-ray flux from WIMP pair annihilation critically depends upon the assumptions one makes on the dark matter profile in the inner portions of the halos. This spread can be extremely large in the case of the nearby Galactic Center, where the dark matter distribution is poorly constrained by observational data.  One is then forced to extrapolate the assumed dark matter profile to very small regions around the center of the Galaxy; the small scale central structure of dark matter halos plays, instead, a less crucial role  when the source is located further away \cite{gammaraysref1}, as in the case of nearby dwarf satellite galaxies \cite{gammaraysref1,draco,Colafrancesco:2006he} or of galaxy clusters \cite{Colafrancesco:2005ji}. A second issue involved in the evaluation of the possibility of detecting a WIMP annihilation signal in gamma-ray data is related to the evaluation of the background. In short, any evaluation of the detectability of a WIMP induced gamma-ray signal must be carefully and properly put in a specific context; comparing the detection perspectives for different astrophysical WIMP annihilation locations can be even more difficult, and full details about the assumptions involved have to be specified.

\begin{figure}[t]
\begin{center}
\includegraphics[width=0.9\textwidth]{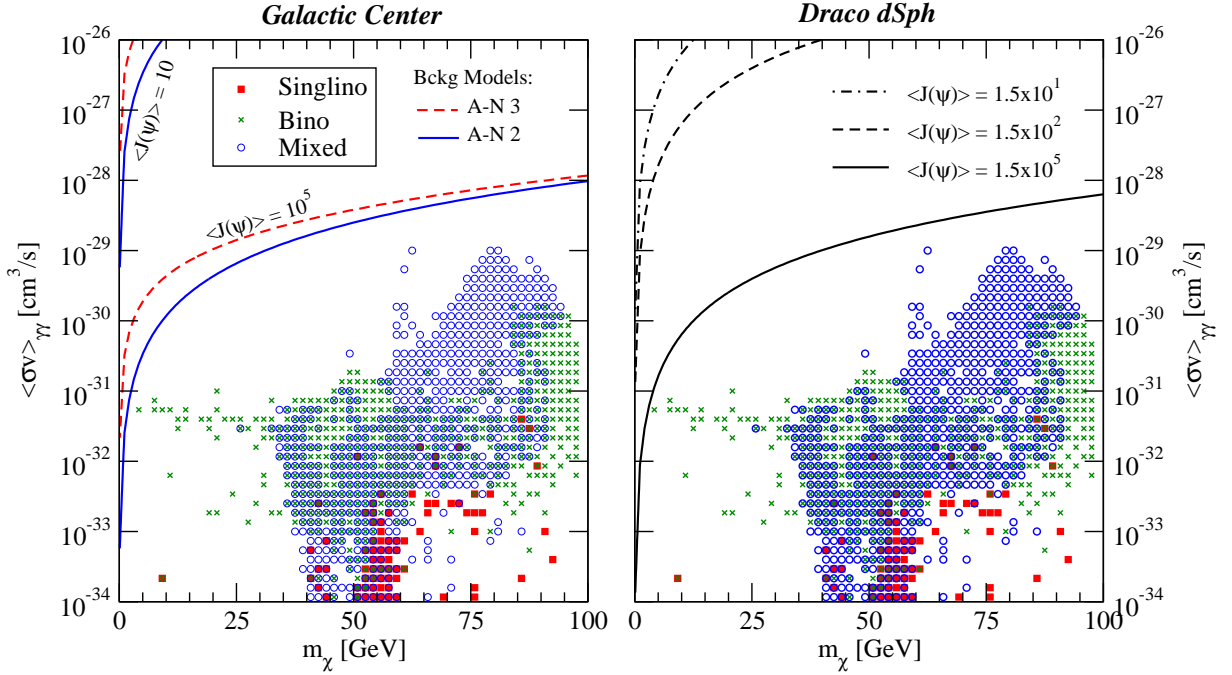}
\end{center}
\caption{Prospects for the detection, with GLAST, of the monochromatic $\gamma\gamma$ gamma-ray line in the Galactic Center region (left) and in the Draco dSph (right). See the text for details. The conventions for the various neutralino types are as in Fig.~\ref{fignmssm}.}
\label{fig:gc_draco}
\end{figure}

In Fig.~\ref{fig:gc_draco} we compare the detection prospects, in the $(m_\chi, \langle\sigma v\rangle_{\gamma\gamma})$ plane (where $\langle\sigma v\rangle_{\gamma\gamma}\equiv\langle\sigma v\rangle\times{\rm BR}(\chi\chi\rightarrow\gamma\gamma)$), of the $\gamma\gamma$ line from neutralino pair-annihilations in the Galactic Center (left) and in the Draco dwarf spheroidal galaxy (right) with the predictions we obtain in our scan over NMSSM models. We consider the sensitivity of GLAST after five years of data taking time $T$, assuming an average angular sensitivity of $\Delta\Omega\simeq9\times10^{-5}$ sr, and an average effective area $A_{\rm eff}$ of 5000 cm${}^2$ \cite{glastsens}. We consider a putative energy bin centered around the location of the gamma-ray line, $E_\gamma=m_\chi$, and as wide as the expected energy resolution of GLAST, $\Delta E/E\simeq 0.1$. Namely, we consider the energy interval
\begin{equation}
(\Delta E)_{m_\chi}\equiv m_\chi/1.05\lesssim E_\gamma\lesssim m_\chi\times1.05.
\end{equation} 
Given a background with a differential flux 
\begin{equation}
\frac{{\rm d}\phi_b}{{\rm d}E}\simeq\phi_0\left(\frac{E}{1\ {\rm GeV}}\right)^{-\gamma}
\end{equation}
we obtain, over the considered energy range a total background flux of 
\begin{equation}
\phi_b=\frac{\phi_0}{\gamma-1}\left(\left(\frac{m_\chi/1.05}{1\ {\rm GeV}}\right)^{1-\gamma}-\left(\frac{m_\chi\times1.05}{1\ {\rm GeV}}\right)^{1-\gamma}\right)\  {\rm cm}^{-2}{\rm s}^{-1}.
\end{equation}
The signal flux from the monochromatic line is instead given by
\begin{equation}
\phi_s=1.87\times 10^{-11}\left(\frac{2\times\langle\sigma v\rangle_{\gamma\gamma}}{10^{-29}\ {\rm cm}^3{\rm s}^{-1}}\right)\left(\frac{10\ {\rm GeV}}{m_\chi}\right)^2\cdot \overline J(\psi,\Delta\Omega)\cdot\Delta\Omega\ {\rm cm}^{-2}{\rm s}^{-1}.
\end{equation}
%------------------------------------------------------------------------------
%------------------------------------------------------------------------------

%------------------------------------------------------------------------------
\begin{figure}[t]
\begin{center}
\includegraphics[width=0.9\textwidth]{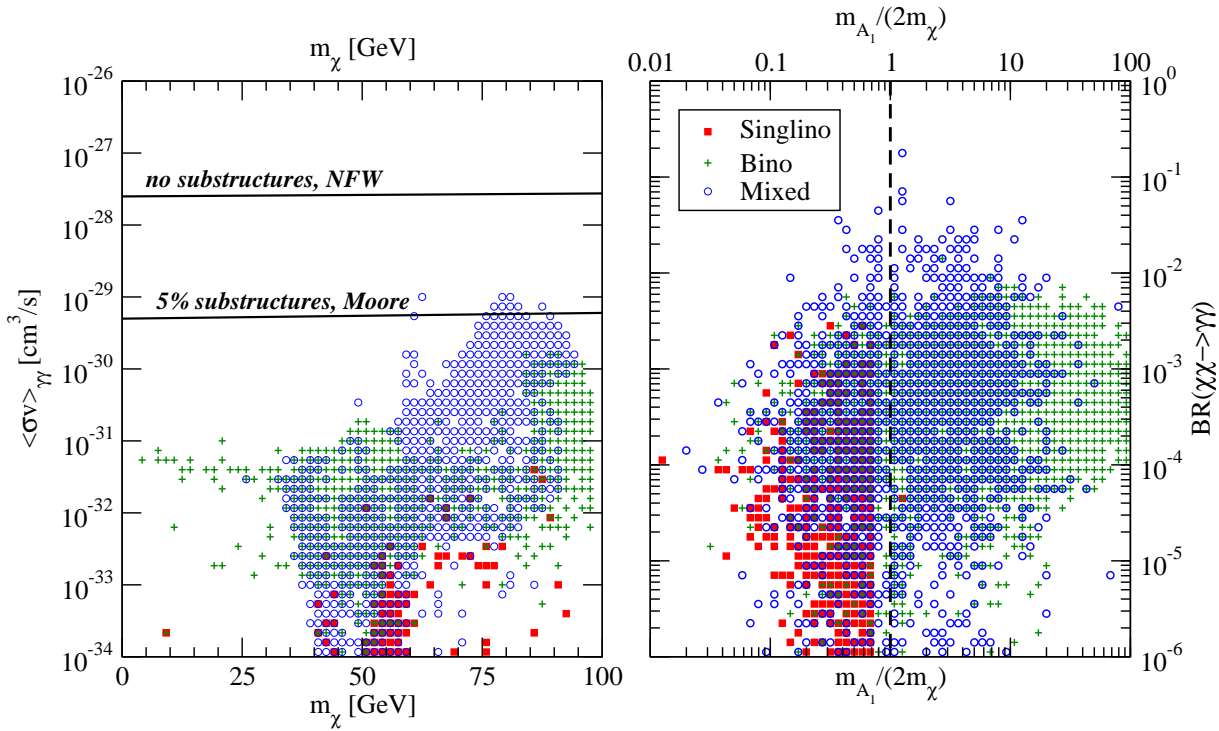}
\end{center}
\caption{Prospects for the detection, with GLAST, of the monochromatic $\gamma\gamma$ gamma-ray line in the extra-galactic gamma-ray background. See the text for details. The conventions for the various neutralino types are as in Fig.~\ref{fignmssm}.}
\label{fig:gammas}
\end{figure}
%------------------------------------------------------------------------------
In the formula above, we defined the dimensionless quantity
\begin{equation}
\overline J(\psi,\Delta\Omega)\equiv\frac{1}{\Delta\Omega}\int_{\Delta\Omega}{\rm d}\Omega\frac{1}{8.5\ {\rm kpc}}\left(\frac{1}{0.3\ {\rm GeV}/{\rm cm}^3}\right)^2\int_{\rm line\ of\ sight}\rho_{\rm DM}^2(l){\rm d}l(\psi).
\end{equation}
We define a signal as ``{\em detectable}'' provided the number of signal events in the considered energy bin $N_s$ is larger than 5, {\em and} the following $5-\sigma$ significance condition is fulfilled:
\begin{equation}\label{eq:criterion}
\phi_s\sqrt{\frac{A_{\rm eff}\cdot T}{\phi_b+\phi_s}}>5.
\end{equation}
Evaluating the gamma-ray background in the Galactic Center is certainly a non trivial task. Since the EGRET data from the Galactic Center likely include a gamma-ray source with a significant offset with respect to the actual Galactic Center \cite{Hooper:2002fx}, we shall consider here the data from the HESS collaboration \cite{Aharonian:2004wa}, which feature a much better angular resolution. The HESS data from the Galactic Center region indicate a steady power law gamma-ray source with a spectrum ${\rm d}N_\gamma/{\rm d}E\propto E_\gamma^{-2.2}$ extending over a range of gamma-ray energies of almost two orders of magnitude \cite{Aharonian:2004wa}. The flux at low energies, $E_\gamma\simeq200$ GeV, is limited by the experimental energy threshold. Extrapolating down to the energies of interest here (a few GeV up to 100 GeV) involves invoking a particular nature for the mechanism responsible for the gamma-ray production. Following \cite{Zaharijas:2006qb}, we consider two extreme choices for the background extrapolation at lower energies, namely the models number 2 and 3 of Aharonian and Neronov, Ref.~\cite{Aharonian:2004jr}, (we shall indicate hereafter the two models as A-N2 and A-N3), respectively giving the smallest and the largest extrapolated background levels among those considered in \cite{Zaharijas:2006qb}. Model A-N2 invokes inelastic proton-proton collisions of multi-TeV protons in the central super-massive black-hole accretion disk, while model A-N3 results from curvature and inverse Compton radiation. We assume $\phi_0\simeq1\times10^{-9}\ {\rm cm}^2{\rm s}^{-1}{\rm GeV}^{-1}$ and $\gamma=2.0$ for model A-N2, while $\phi_0\simeq3\times10^{-7}\ {\rm cm}^2{\rm s}^{-1}{\rm GeV}^{-1}$ and $\gamma=2.75$ for model A-N3.

As far as the values of $\overline J(0,\Delta\Omega)$ are concerned, we consider the range given by the extrapolation of the two halo models considered above (the Burkert and the adiabatically contracted N03 profiles), giving, roughly, $\overline J\simeq 10$ and $\overline J\simeq 10^5$. The left panel of Fig.~\ref{fig:gc_draco} illustrates our results. All the sensitivity lines correspond to the criterion given in (\ref{eq:criterion}), which we find to be always more stringent than the $N_s>5$ requirement. The solid blue lines correspond to the A-N2 background model, while the A-N3 background is assumed for the red dashed lines. Our results show that even assuming a very optimistic dark matter profile, the likelihood of obtaining a significant gamma-ray line detection from the Galactic Center is rather low; at the best, a weak excess can be detected either with very low mass neutralinos, or with mixed neutralino-singlino models with a mass $m_\chi\sim60\div80$ GeV.

In the case of Draco, the estimated background is only given by the diffuse gamma-ray background, which we parameterize with $\phi_0\simeq6.3\times10^{-11}\ {\rm cm}^2{\rm s}^{-1}{\rm GeV}^{-1}$ and $\gamma=2.1$. We follow the results of Ref.~\cite{Colafrancesco:2006he} as far as the estimate of $\overline J$ are concerned; conservatively, a range of viable halo profiles for Draco gives $10\lesssim\overline J\lesssim 100$. Taking into account the possibility of a central super-massive black-hole and the subsequent adiabatic of dark matter in a central ``{\em spike}'' \cite{Colafrancesco:2006he} can greatly enhance the viable values of $\overline J$, up to the level of $10^5$, a value we assume for the black solid line. As for the Galactic Center, the prospects of cleanly detecting a gamma-ray line from the direction of Draco do not seem particularly exciting, although, again, some models might in principle, and very optimistically, give some evidence of an energy-localized gamma-ray excess.

The monochromatic WIMP pair annihilation is also constrained by the contribution that annihilations occurring in any dark matter halo and at all redshifts give to the extragalactic gamma-ray radiation \cite{Bergstrom:2001jj,Ullio:2002pj}. We refer the reader to the thorough discussion given in Ref.~\cite{Ullio:2002pj}, and we make use here of the constraints, on the $m_\chi, \langle\sigma v\rangle_{\gamma\gamma}$ plane derived in Fig.~15 of the same study. In particular, we report in Fig.~\ref{fig:gammas}, left, the sensitivity, on the above mentioned plane, expected from GLAST, under the two extreme scenarios for the halo profiles and the presence of dark matter substructures outlined in \cite{Ullio:2002pj}. The upper curve refers to halos modeled by a NFW profile \cite{NFW}, no substructures and concentration parameters inferred from the Bullock et al. model \cite{bullock}, while the lower curve assumes the (cuspier) Moore et al. profile \cite{moore}, with 5\% of the halo mass in substructures with concentration parametes 4 times than that estimated with the Bullok et al. model. In the most generous scenario, a few mixed singlino-bino models can give rise to a detectable signal at GLAST, although more conservative assumptions leave small space for any hope of detecting any signature at all in the extra-galactic gamma-ray data.

Even though the prospects for the detection of the monochromatic line do not look particularly promising here, we wish to point out that the branching fractions we find, and the absolute values of $\langle\sigma v\rangle_{\gamma\gamma}$ are, typically, larger than in the MSSM. We devote App.~\ref{sec:grapp} to a detailed discussion of this point, but we wish here to emphasize the main reason why the NMSSM rate for the process $\chi\chi\rightarrow\gamma\gamma$ is expected to be more significant than in the MSSM. The potentially light extra CP-odd gauge boson gives rise to the extra contributions shown in Fig.~\ref{fig:feyngg}; the size of this contribution, generically, depends upon whether the annihilation proceeds close to the $s$-channel resonance ($m_\chi\simeq m_{a_1}/2$). We illustrate the effect of the extra NMSSM diagrams in Fig.~\ref{fig:gammas}, right, where we show the size of ${\rm BR(\chi\chi\rightarrow\gamma\gamma)}$ as a function of the ratio $m_{a_1}/(2m_\chi)$. As evident from the figure, the largest branching ratios occur when $m_{a_1}/(2m_\chi)\sim1$, and they are more than a couple of orders of magnitude larger than the MSSM limit (bino-like neutralinos and $m_{a_1}/(2m_\chi)\gg1$).

\section{Conclusion}

Gauge singlet extensions of the Higgs sector of the minimal supersymmetric Standard Model provide well motivated theoretical and phenomenological laboratories. Besides offering an elegant solution to the supersymmetric $\mu$ problem, they provide a viable way out of the difficulties connected to encompassing a mechanism of electroweak baryogenesis in the MSSM. In the present analysis, we focused on one specific such extension, the so-called NMSSM, and investigated, for the first time, the prospects for neutralino dark matter indirect detection. 

The present study is motivated by two basic observations: first, in the NMSSM, unlike the MSSM, the lightest neutralino can be naturally very light, as a result of the possibility of it annihilating through a potentially very light extra CP-odd, mostly singlet-like Higgs boson; second, the extended Higgs sector leads to extra diagrams in the loop-amplitude relevant for the pair annihilation of neutralinos in photon or gluon pairs. An enhancement of the monochromatic $\chi\chi\rightarrow\gamma\gamma$ gamma-ray line is therefore generically expected within the NMSSM, as opposed to the minimal supersymmetric setup.

We found that the rate of neutrinos produced by the annihilation of neutralino dark matter particles captured inside the Earth and the Sun is in general large in the NMSSM; unlike the MSSM, we found that most models give a larger signal from annihilations in the core of the Earth rather than in the Sun, at a level which can, in certain cases, be constrained by current available data from SuperKamiokande and MACRO.  This is presumably due in part to additional low velocity contributions to the local neutralino density in the region of the Earth which can result for light neutralinos. Future neutrino telescopes with increased sensitivity
for low neutrino energies will be able to probe a sizable part of the parameter
space by looking at signals from both the Earth and the Sun.

The dynamics of the Sun could also be modified due to energy transport
by neutralinos. Our estimates show that, especially bino-like, 
neutralinos below $m_\chi
\lesssim 30\ \gev$ might contribute a significant fraction of the total
solar luminosity. More detailed studies, using self-consistent
solar models, could unveil large enough modifications 
on the sound speed or on the boron neutrino flux to significantly disfavor light neutralino scenarios.

We showed that within the NMSSM the expected antimatter yield from neutralino pair annihilations in the galactic halo can be sizable, although the absolute normalization of the flux depends on specific assumptions about the dark matter halo profile. In particular, we found that signals at low-energy antideuteron search experiments such as GAPS, and at space-based antimatter search experiments such as PAMELA, are expected, though not guaranteed, for very light neutralinos ($m_\chi\lesssim 20$ GeV) or for intermediate mass mixed singlino-bino neutralinos ($60\lesssim m_\chi\lesssim 90$ GeV).

We worked out for the first time the loop-induced pair annihilation cross section for NMSSM neutralinos into two photons and two gluons, pointing out that the expected branching ratio, with respect to tree-level neutralino pair annihilation into other Standard Model particles, is typically large, especially when compared to the MSSM case. The reason for this enhancement is traced back to diagrams which are resonant when $2 m_\chi\simeq m_{a_1}$, the latter quantity indicating the mass of the lightest, extra CP-odd Higgs boson.

We analyzed in detail the prospects for the detection of the monochromatic gamma-ray line resulting from $\chi\chi\rightarrow\gamma\gamma$ annihilation processes in the Galactic Center, in a nearby dwarf spheroidal Galaxy (Draco) and the coherent effect of annihilations in any dark matter halo contributing to the extra-galactic gamma-ray radiation. We pointed out that most models are not expected to give any detectable signal at GLAST, although this detection channel looks significantly more promising than in the usual MSSM setup.

Finally, with the purpose of making the present study a useful and complete starting point for future research in the field, and in order to sort out and clarify some notational ambiguities and inconsistencies, we collect in the Appendix the details of the one-loop computation of the $\chi\chi\rightarrow\gamma\gamma,\ gg$ amplitudes and other quantities relevant for the estimate of indirect detection rates.

\begin{acknowledgments}

We gratefully acknowledge useful conversations with John Beacom, Marco Cirelli,
Ulrich Ellwanger, Cyril Hugonie, Bob McElrath, Alexander Pukhov and
Miguel Angel Sanchis-Lozano. 

FF and LMK are supported in part by grants from the DOE and NSF at Case Western Reserve University.  SP is supported in part by DOE grants DE-FG03-92-ER40701 and
FG02-05ER41361 and NASA NNG05GF69G.   

\end{acknowledgments}

\appendix

\section{Neutralino annihilation channels}
\label{appannihilation}
\subsection{Tree level processes}

Analytic expressions for the \nmssm-like neutralino annihilation into two 
particles at tree level can be found in~\cite{Stephan:1997ds}\footnote{The
notation in~\cite{Stephan:1997ds} follows the usual \mssm\ practice of
labeling the CP-even Higgs mass eigenstates as $H_1$ and $H_2$. To make
contact with our conventions one has to switch the indices $1 \leftrightarrow
2$ in the scalar Higgs matrix $S_{ij}$, and $3 \leftrightarrow 4$ in
the neutralino matrix $N_{ij}$. Furthermore, $\la$ and $\k$ in the
superpotential have the opposite sign as ours.}.

For the study of indirect detection, only the nonzero terms in the 
limit $v \rightarrow 0$ need to be taken into account, 
which restricts the relevant processes to those with a CP-odd final state
that have a non-vanishing $S$-wave amplitude.

With respect to the \mssm, the main differences are:
\begin{itemize}
\item An additional scalar Higgs, $h_3$, exchange in the $s$ channel 
contributes to the annihilation to $W^+W-$ and $ZZ$.
\item A fifth neutralino is exchanged in the $t$ and $u$ channels for the
$ZZ$ process.
\item There is an extra $Z$ -- scalar Higgs final state: $Z h_3$. Two
Higgs pseudo-scalars instead of one, and five neutralinos contribute to these
reactions.
\item For the $W^-H^+$ final state, one has to take into account the
contribution of the extra $h_3$ and $a_2$ Higgses.
\item There are five additional final states with a scalar and a pseudoscalar
Higgs. Diagrams $\wt{\chi_5^0}$ have to be considered.
\item Finally, for the $f-\bar{f}$ final state, one has to include
the exchange of the additional Higgses $h_3$ and $a_2$.
\end{itemize}

On top of that, the different couplings have contributions proportional to
$\la$ and $\k$ not in the \mssm~\cite{Ellwanger:2005dv}. 
We used the $v$-independent part of the $S$-wave terms
in~\cite{Stephan:1997ds} for our predictions in Section~\ref{sec:id}.

\subsection{One loop processes}

\subsubsection{Neutralino annihilation into two photons}\label{sec:grapp}

Some radiative processes, even if loop suppressed, are of interest
for dark matter detection. The annihilation to two photons,
$\lsp \lsp \rightarrow \gamma \gamma$, has a characteristic monochromatic
signature at $E_\gamma = m_\lsp/2$. This allows a clear distinction from
all astrophysical backgrounds, unlike the continuum spectrum produced in
tree level processes.

In the context of the \mssm, a full one-loop calculation was performed
in~\cite{Bergstrom:1997fh,Bern:1997ng}. We computed the cross-section for this
process in the \nmssm\ by adapting the results
of~\cite{Bergstrom:1997fh}\footnote{The expressions for the \mssm\ 
were implemented in the code \ds~\cite{Gondolo:2004sc}. 
We have adapted the relevant subroutines to \nmhdecay .}.

\begin{figure}
\begin{center}
\begin{tabular}{c@{\hspace{1.5cm}}c}
  \includegraphics[width=0.25\textwidth]{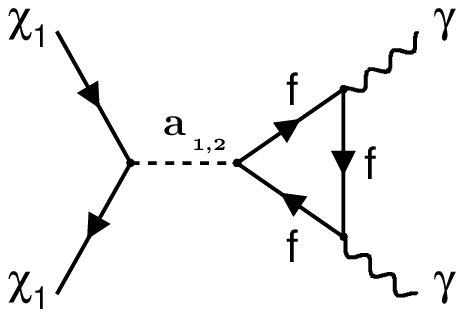}&
  \includegraphics[width=0.25\textwidth]{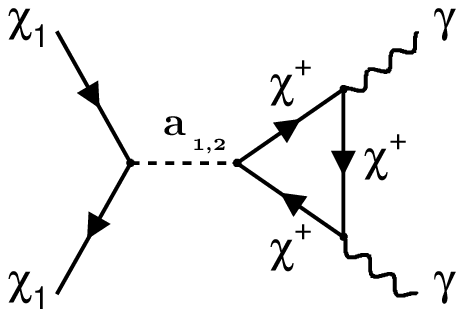}
\end{tabular}
\end{center}
\caption{Additional diagrams, due to the second CP-odd Higgs boson,
for the process
$\lsp \lsp \rightarrow \gamma \gamma$ in the \nmssm.
The contribution of $a_1$ corresponds to Fig.~1.d (left)
and 2.d (right) in~\cite{Bergstrom:1997fh} for the \mssm.}
\label{fig:feyngg}
\end{figure}

Apart from the dependence on $\la$ and $\k$ of the \nmssm\ couplings,
we need to compute two additional diagrams, shown in Fig.~\ref{fig:feyngg},
due to the presence of a second pseudoscalar Higgs boson, $a_2$.

Four types of Feynman diagrams, contributing to the two
photon annihilation amplitude, were identified in~\cite{Bergstrom:1997fh}.
Let us discuss their computation in the \nmssm\ in turn:

\paragraph{Diagrams 1.a-1.d}
For the fermion--sfermion loop diagrams, we need to duplicate the 
CP-odd Higgs terms (Fig.~\ref{fig:feyngg}) and substitute the \nmssm\ couplings
in $S_{f\bar{f}}$, $D_{f\bar{f}}$, $G_{Zf}$ and $G_{a_i f}$
of Eqs.~(7) and~(8) in~\cite{Bergstrom:1997fh}.

For up-type quarks, let us define:
\bea
g_{ll}&=&\frac{-g_2 m_q N_{13}}{\sqrt{2} m_W \sin \beta} \nn \\
g_{rl}&=&\mp \frac{g_2 N_{12}+(\pm 2 e_q-1) g_y N_{11}/3}{\sqrt{2}} \nn \\
g_{lr}&=& \sqrt{2} g_y N_{11} e_q \nn \\
g_{rr}&=& g_{ll}
\label{glr}
\eea
\noi where $g_2$ is the electroweak coupling constant,
$g_y=g_2 \tan \theta_W$, $e_q$ is the quark charge and $m_q$ its mass.
For down quarks and leptons we need to replace $N_{13}\rightarrow
N_{14}$ and $\sin \beta \rightarrow \cos \beta$ and use the lower
sign for $g_{rl}$.
\noi Then, with:
\bea
g_1&=& g_{ll} \cos \theta_{\wt{q}} + g_{lr} \sin \theta_{\wt{q}} \nn \\
g_2&=& - g_{rl} \sin \theta_{\wt{q}} + g_{rr} \cos \theta_{\wt{q}}
\label{g12}
\eea
\noi where $\wt{q}$ is the squark mixing angle which is taken to be
$\wt{q}=0$ ({\em i.e.} no mixing) for the first two families in \nmhdecay.

\noi we have:
\bea
S_{f\bar{f}}&=& \frac{g_1^2+g_2^2}{2} \nn \\
D_{f\bar{f}}&=& g_1 g_2
\eea

The $Z \lsp \lsp$ coupling reads:
\beq
G_{Zf}=\frac{g_2^2 T_{3f}}{\cos^2 \theta_w} \left(N_{13}^2-N_{14}^2\right)
\label{gzf}
\eeq
\noi where the weak isospin, $T_{3f}$ is $+ 1/2$ for up-type quarks and
$-1/2$ for down quarks and leptons.

Finally, using the $a_i \lsp \lsp$ coupling from~\cite{Ellwanger:2005dv}:
\bea
g_{a_a \widetilde{\chi}_i^0 \widetilde{\chi}_j^0}&=&
\frac{\la}{\sqrt{2}} (P_{a1}
\Pi_{ij}^{45} + P_{a2} \Pi_{ij}^{35} + P_{a3} \Pi_{ij}^{34}) - \sqrt{2}
\k P_{a3} N_{i5} N_{j5} \nn \\ 
& &  + \frac{g_y}{2} (P_{a1} \Pi_{ij}^{13} - P_{a2}
\Pi_{ij}^{14}) - \frac{g_2}{2} (P_{a1} \Pi_{ij}^{23} - P_{a2}
\Pi_{ij}^{24}),
\label{achichi}
\eea
\noi where $\Pi_{ij}^{ab} = N_{ia}N_{jb}+N_{ib}N_{ja}$, we obtain:
\beq
G_{a_i f}=-2 g_{a_i \lsp \lsp} \frac{m_q g_2 P_{i1}}{m_w \sin \beta}
\label{gaf}
\eeq
\noi Changing
$\sin \beta \rightarrow \cos \beta$ and $P_{i1} \rightarrow P_{i2}$ in
Eq.~(\ref{gaf}), leads to the corresponding expression for down-type
quarks and leptons.

\paragraph{Diagrams 2.a-2.d}

For the chargino--Higgs loop diagrams need also to take into account the
additional contribution of $a_2$ and use the expressions below
for $S_{\chi H}$,
$D_{\chi H}$, $G_{Z \chi}$ and $G_{a_i \chi}$ in Eq.~(9)
of~\cite{Bergstrom:1997fh}.

Taking the $H^+ \lsp \chi^-$ coupling from~\cite{Ellwanger:2005dv}:
\bea
g_{H^+ \chi^-_i \chi^0_j}&=&\la\cos\b U_{i2} N_{j5} -
\frac{\sin\b}{\sqrt{2}} U_{i2} (g_y N_{j1} + g_2 N_{j2}) + g_2 \sin\b
U_{i1} N_{j4} \nn \\
g_{H^- \chi^+_i \chi^0_j}&=& \la\sin\b V_{i2} N_{j5} +
\frac{\cos\b}{\sqrt{2}} V_{i2} (g_y N_{j1} + g_2 N_{j2}) + g_2 \cos\b
V_{i1} N_{j3},
\label{hcchichar}
\eea
\noi where $U$ and $V$ are the chargino mass matrices.
we get:
\bea
S_{\chi H}&=& \frac{g_{H^+ \chi^-_i \lsp}^2
+g_{H^- \chi^+_i \lsp}^2}{2} \nn \\
D_{\chi H}&=& g_{H^+ \chi^-_i \lsp} g_{H^- \chi^+_i \lsp}.
\eea

The $Z$ exchange diagrams require:
\beq
G_{Z \chi}=\frac{g_2^2}{\cos^2 \theta_w} \left(V_{i1}^2+\frac{V_{i2}^2}{2}-
U_{i1}^2-\frac{U_{i2}^2}{2}\right)  \left(N_{13}^2-N_{14}^2\right)
\eeq

With Eq.~(\ref{achichi}) and the $a_a \chi^+_i \chi^-_j$ coupling:
\beq
g_{a_a \chi^+_i \chi^-_j}=\frac{\la}{\sqrt{2}} P_{a3} U_{i2}
V_{j2} - \frac{g_2}{\sqrt{2}} (P_{a1} U_{i1} V_{j2} + P_{a2} U_{i2}
V_{j1}),
\label{acharchar}
\eeq
we have:
\beq
G_{a_i \chi}=-4 g_{a_i \chi^+_j \chi^-_j}  g_{a_i \lsp \lsp}
\eeq

\paragraph{Diagrams 3.a-3.c}

The $W^+ \lsp \chi^-_j$ couplings are:
\bea
g^L_{W1i}&=&g_2 \left(-N_{13} V_{i2}/\sqrt{2}
+N_{12}V_{i2}\right)  \nonumber \\
g^R_{W1i}&=&g_2 \left(N_{14} U_{i2}/\sqrt{2}
+N_{12} U_{i2}\right) 
\eea
\noi which we can substitute in $S_{\chi W}$ and $D_{\chi W}$ of
Eqs.~(11) and~(12) in~\cite{Bergstrom:1997fh} to obtain the chargino-W
loop contribution.

\paragraph{Diagrams 4.a-4.b}

The unphysical Higgs boson is orthogonal to the charged Higgs and we can
derive its required coupling to neutralinos and charginos by adapting
those of the charged Higgs in~\cite{Ellwanger:2005dv}:

\bea
g_{G^+ \chi^-_i \chi^0_j} & = & \la\sin\b U_{i2} N_{j5} -
\frac{\cos\b}{\sqrt{2}} U_{i2} (g_y N_{j1} + g_2 N_{j2}) + g_2 \cos\b
U_{i1} N_{j4} \nonumber \\
g_{G^- \chi^+_i \chi^0_j} & = & \la\cos\b V_{i2} N_{j5} +
\frac{\sin\b}{\sqrt{2}} V_{i2} (g_y N_{j1} + g_2 N_{j2}) + g_2 \sin \b
V_{i1} N_{j3}. 
\label{goldchi}
\eea
Then, in Eq.~(13) of~\cite{Bergstrom:1997fh} we need to input:
\bea
S_{\chi G}&=& \frac{g_{G^+ \chi^-_i \lsp}^2
+g_{G^- \chi^+_i \lsp}^2}{2} \nn \\
D_{\chi G}&=& g_{G^+ \chi^-_i \lsp} g_{G^- \chi^+_i \lsp}.
\eea

In the expressions above, we have not taken into account that \nmhdecay\ uses
a real neutralino and chargino mass matrix, whereas the expressions
in~\cite{Bergstrom:1997fh} assume that diagonalization in the neutralino and
chargino sectors is performed using a complex $N$, $U$ and $V$, so that
$m_\lsp$ and the chargino masses are always positive.

To correct for this fact we need to multiply by $\epsilon\equiv{\rm sign}
\left(m_\lsp\right)$ all instances of $N^*$ in~\cite{Bergstrom:1997fh}
for a vertex in which the
neutralino is annihilated~\cite{Gunion:1984yn} and a similar change of
sign needs to be done in $V$. Details for each vertex
can be found, for the \mssm, in~\cite{Edsjo:1997hp}. For the two photon
amplitude computation, this prescription amounts to multiplying by $\epsilon$
($\eta\equiv {\rm sign} (m_\lsp m_{\chi^+})$)
the $D$ terms in diagrams of type 1 (2, 3 and 4).

\begin{figure}[hbt]
\begin{center}
\begin{tabular}{c}
\includegraphics[width=0.5\textwidth]{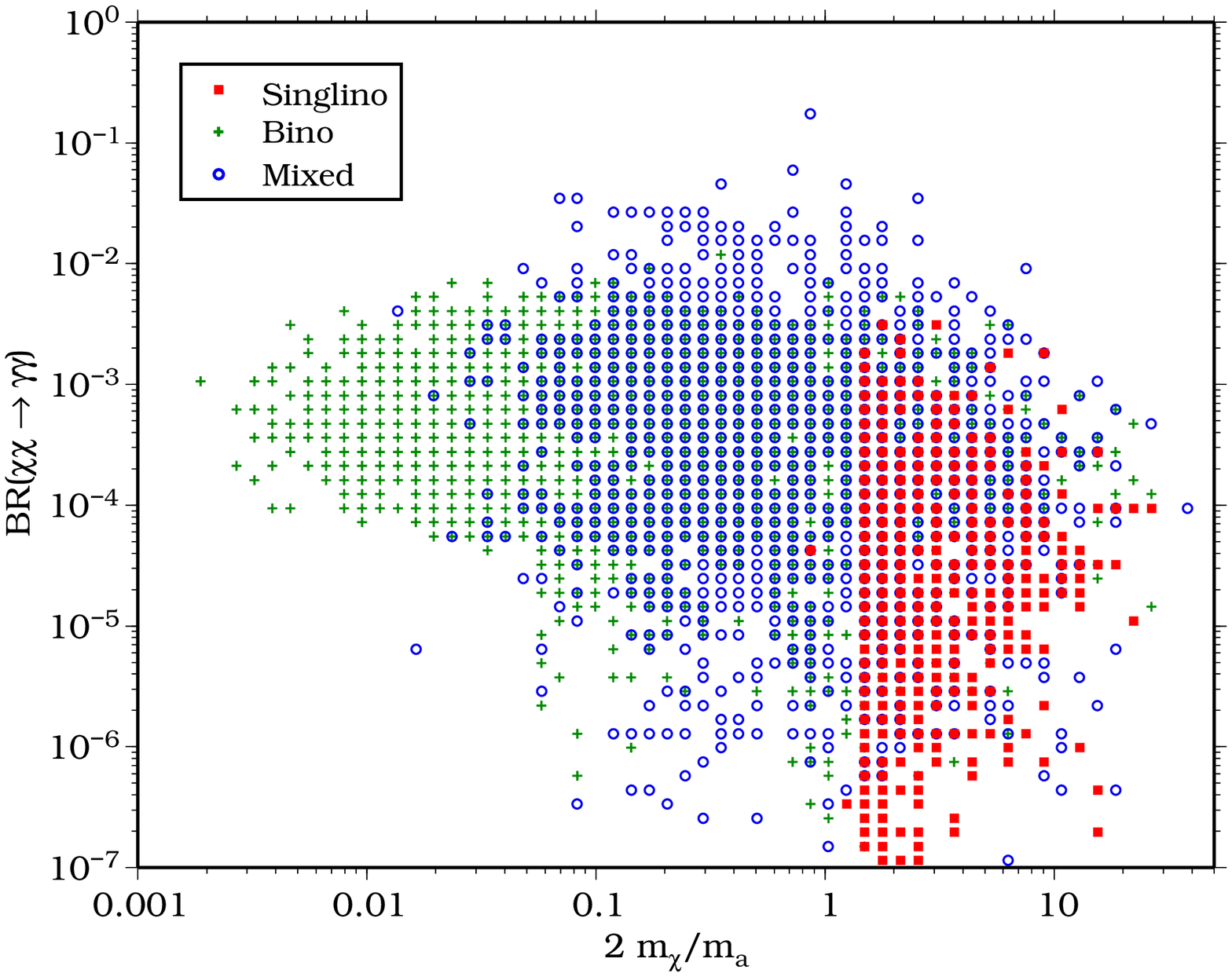}\\
\begin{tabular}{c@{\hspace{1cm}}c}
  \includegraphics[width=0.45\textwidth]{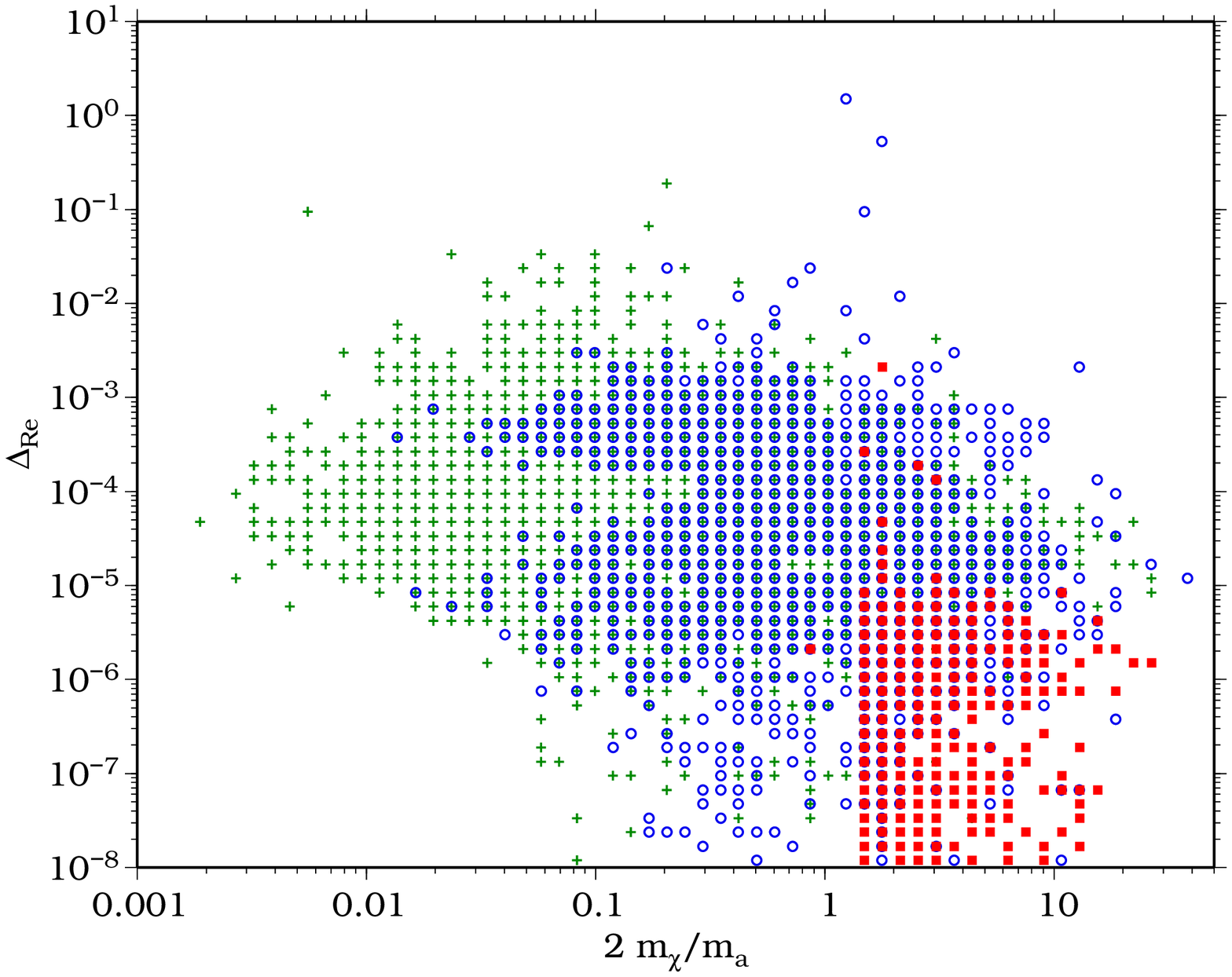}&
  \includegraphics[width=0.45\textwidth]{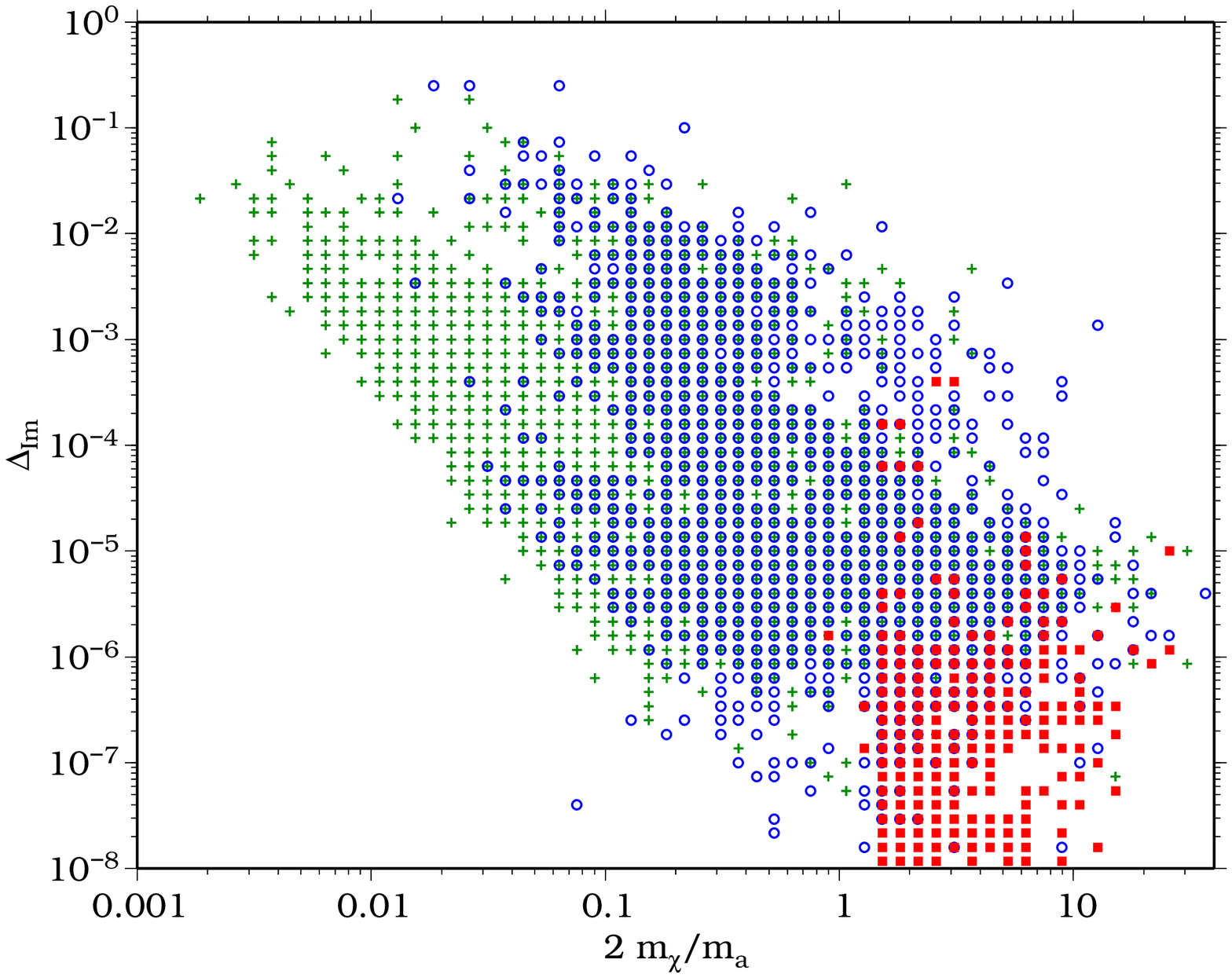}
\end{tabular}
\end{tabular}
\end{center}
\caption{Branching ratio for the annihilation to $\gamma \gamma$ (up) and
contribution of the CP-odd Higgs
exchange diagrams to the real (left) and imaginary (right) 
part of the amplitude.}
\label{fig:ggh3}
\end{figure}

The presence of extra light CP-odd Higgses, can enhance the cross
section for this process. In Fig.~(\ref{fig:ggh3}), we show the branching
ratio for the process $\lsp \lsp \rightarrow \gamma \gamma$ together with
the contribution from the diagrams in Fig.~\ref{fig:feyngg}.

The branching ratio peaks for neutralino masses $2 m_\lsp \sim m_{a_1}$, 
more so, for singlino and mixed-like neutralinos\footnote{Note the absence
of singlino-like neutralinos with $m_\lsp \leq  m_{a_1}/2$.}. The contribution
of the CP-odd Higgs diagrams, Fig.~\ref{fig:feyngg}, 
to the total Feynman amplitude, ${\cal A}$, is displayed in the lower
panels by the quantity 
\beq
\Delta=\frac{1}{\left| 1-\frac{{\cal A}_{a_i}}{{\cal A}}\right|}.
\eeq
\noi Larger values of $\Delta$, corresponding to larger relative contributions
of the CP-odd diagrams to the total amplitude, also cluster for those
values where the branching ratio is larger. 
%Even for bino-light neutralino,
%when the larger branching ratios lie in the region $2 m_\lsp \ll m_{a_1}$, 
%which is where $\Delta$ is larger.

The light CP-odd Higgses, together with the additional singlino component,
lead, thus, to an enhancement of the $\gamma \gamma$ annihilation channel
in the \nmssm\ compared to the \mssm.

\subsubsection{Annihilation into two gluons}

The cross section for this process~\cite{Drees:1993bh}
can be obtained at once from the two photon
channel computed in the previous section. In order to do so, we need to
consider only the diagrams of type 1 for quarks, with no contribution 
from leptons. The electric charge is substituted by $e_q^2 \rightarrow 1$,
and the color sum average is performed by $\alpha_{em}^2 \rightarrow
2 \alpha_{s}^2$ in the final expression for $\sigma v$~\cite{Bergstrom:1997fh}.

%\subsubsection{Annihilation into $Z \gamma$}

\section{Elastic scattering cross sections}
\label{appelastic}
We review here the computation of the neutralino-nucleon elastic cross-section,
which is used to predict direct detection rates and, in the context of
indirect detection, determines the capture rate of neutralinos in the Sun 
or in the Earth.

The basic ingredient for the neutralino-nucleon cross-section is the
individual neutralino-quark cross-section, which for the \mssm\ can be found 
in~\cite{Drees:1993bu}. The $\chi-q$ process has been studied before 
in the context of the \nmssm. First
in~\cite{Bednyakov:1998is}, where both spin-independent and spin-dependent
contributions were computed. More recently the problem was revisited
in~\cite{Cerdeno:2004xw}, were only the spin-independent part was considered
and a mistake in~\cite{Bednyakov:1998is} was corrected. The authors
in~\cite{Gunion:2005rw} approximated the spin-independent interaction
by assuming that the $t$--channel exchange of CP-even Higgses dominates. For
our predictions in Sec.~\ref{sec:id}, we re-derived the $\lsp-q$ cross-section
for the \nmssm\ by extending the \mssm\ calculation~\cite{Drees:1993bu}.

%\subsection{Neutralino--quark elastic scattering cross section}

The low energy $\chi-q$ effective lagrangian can be written as:
\beq
{\cal L}_{eff}= d_q \bar{\wt{\chi}} \gamma^{\mu}\gamma_5 \wt{\chi}
\bar{q}\gamma_{\mu}\gamma_5 q+
f_q \bar{\wt{\chi}} \wt{\chi} \bar{q} q
\label{efflag}
\eeq
where only contributions that don't vanish when $v\rightarrow 0$ have been
written. The first term describes the spin-dependent contribution and 
the second one the spin-independent one.

As in the \mssm\ we have two types of diagrams contributing to the 
spin-dependent interaction in the \nmssm, $Z$ exchange and squark 
exchange. For the spin-independent one, CP-even Higgs exchange and squark
exchange contribute to $f_q$.

Following~\cite{Drees:1993bu}, let us define:
\bea
X&=&-\sqrt{2}\left[g_2 T_{3f}N_{12}-g_y \left(T_{3f}-e_q\right)N_{11}\right]
 \nonumber \\
Y&=&\sqrt{2} g_y e_q N_{11}\nonumber \\
Z_{up}&=&-\frac{g_2 m_q N_{13}}{\sqrt{2} \sin \beta m_w} \quad \quad
Z_{down}=-\frac{g_2 m_q N_{14}}{\sqrt{2} \cos \beta m_w}.
\label{xyz}
\eea
\noi Then the couplings involving the lightest squark can be written as:
\bea
a_{\wt{q}_1}&=&\frac{1}{2}\left[ \cos \theta_q (X+Z)+\sin \theta_q (Y+Z)\right]
\nonumber \\
b_{\wt{q}_1}&=&\frac{1}{2}\left[ \cos \theta_q (X-Z)+
\sin \theta_q (Z-Y)\right],
\label{absquark}
\eea
and the corresponding equations for the heavier eigenstate, $j=2$, are
found taking $\sin \theta_q \rightarrow  \cos \theta_q $ and
$\cos \theta_q \rightarrow - \sin \theta_q$.

With that, the spin-dependent $\lsp-q$ interaction is given by:
\beq
d_q =\frac{1}{4} \sum_{j=1}^2{\frac{a^2_{\wt{q}_j}+b^2_{\wt{q}_j}}
{m^2_{\wt{q}_j}-\left(m_\chi+m_q \right)^2}}
-\frac{g_2^2}{4 m_w^2} T_{3q}\frac{1}{2}\left(N_{13}^2-N_{14}^2\right),
\label{alpha2}
\eeq
\noi where the sum runs over the squark eigenstates and the last term
describes the $Z$ exchange contribution.

As for the spin-independent interaction, we have:

\beq
f_q =-\frac{1}{4} \sum_{j=1}^2{\frac{a^2_{\wt{q}_j}-b^2_{\wt{q}_j}}
{m^2_{\wt{q}_j}-\left(m_\chi+m_q \right)^2}}
-m_{q} \sum_{j=1}^3{\frac{g_2 g_{H_j \lsp \lsp} S_{j1}}
{m_{H_j}^2 m_w \sin \beta}}.
\label{alpha3}
\eeq
Note that in the \nmssm\ we have three CP-even Higgses, included in the last
term. For down type quarks, we need to replace in Eq.~(\ref{alpha3}),
$S_{j1}\rightarrow S_{j2}$ and $\sin \beta \rightarrow \cos \beta$. Also,
we need the coupling $q-\wt{q}-\lsp$:
\bea
g_{h_a \chi^0_i \chi^0_j}&=& \frac{\la}{\sqrt{2}} (S_{a1} \Pi_{ij}^{45} +
S_{a2} \Pi_{ij}^{35} + S_{a3} \Pi_{ij}^{34}) - \sqrt{2} \k S_{a3}
N_{i5} N_{j5} \nn \\ & & - \frac{g_1}{2} (S_{a1} \Pi_{ij}^{13} - S_{a2}
\Pi_{ij}^{14}) + \frac{g_2}{2} (S_{a1} \Pi_{ij}^{23} - S_{a2}
\Pi_{ij}^{24}).
\label{hchichi}
\eea

Our expression for the spin-dependent interaction agrees with the
computation in~\cite{Bednyakov:1998is}. As for the spin-independent part,
the authors in~\cite{Cerdeno:2004xw} noted a mistake in the expressions given in~\cite{Bednyakov:1998is}. We agree with their remark, but note
that the $q-\wt{q}-\lsp$ couplings given in~\cite{Cerdeno:2004xw} contain
a mistake, since the couplings for $\wt{q_2}$ cannot be obtained from those
of $\wt{q_1}$ by the usual change $\sin \theta_q \rightarrow  \cos \theta_q$ 
and $\cos \theta_q \rightarrow - \sin \theta_q$. Indeed, the sign
in front of $N^*_{\alpha 2}$ should affect the whole coefficient of
the $\sin \theta_q$ term in their Eq.~(A.11).

Since we have used a real neutralino matrix $N_{ij}$, we should add the
necessary factors of $\epsilon$ in the expressions above. 
We have followed~\cite{Drees:1993bu}, where a real $N_{ij}$ was used, and where it was pointed out that the absolute value should be used in the kinematic factors
appearing in various denominators. However, following the prescription
in~\cite{Gunion:1984yn}, one can check that when $m_\lsp <0$, one should
also take $f_q \rightarrow \epsilon f_q$. This extra sign difference, 
however, does not affect the nucleon cross sections discussed below, for
they depend quadratically on $f_q$ or $d_q$.

%\subsection{Proton-Neutron cross sections}

Once the individual $\lsp-q$ cross sections are determined, we can proceed
to compute the nucleon (proton or neutron) cross sections used in
Sec.~\ref{sec:id}.

The spin-independent nucleon-neutralino elastic cross section is given by:
\beq
\sigma^{si}_{p,n}=\frac{4 m_r^2}{\pi} f_{p,n}^2
\label{sinucl}
\eeq
\noi where
\beq
f_{p,n}=\sum_{q=u,d,s} {f_{Tq}^{p,n} f_{q} \frac{m_{p,n}}{m_q}}+
\frac{2}{27}\left(1-\sum_{q=u,d,s}{f_{Tq}^{p,n} } \right)
\sum_{q=c,b,t} {f_{Tq}^{p,n} f_{q} \frac{m_{p,n}}{m_q}}
\eeq
\noi and for the quark composition of each nucleon, $f_{Tq}^{p,n}$, 
we use the central values found in~\cite{Bertone:2004pz}. In
Eq.~(\ref{sinucl}), the reduced mass 
is $m_r \equiv m_{p,n} m_\chi/\left(m_{p,n} +m_\chi\right)$.

For the axial-vector interactions we need the nucleon spin carried by 
each quark. We use, again, the central values from~\cite{Bertone:2004pz}
to find:
\beq
\sigma^{sd}_{p,n}=\frac{4 m_r^2}{\pi} 3 \left(f_{u} \Delta_u^{p,n}+
f_{d} \Delta_d^{p,n}+f_{s} \Delta_s^{p,n}\right)^2.
\eeq

In Sec.~\ref{solarbounds}, scalar cross sections with nuclei are used.
Since the values at zero momentum transfer are good enough for the task
at hand, we compute them as:

\beq
\sigma_i = \frac{4 m_r^2}{\pi} \left[ Z f_p+(A-Z) f_n \right]^2,
\label{sigmanucl}
\eeq
where $Z$ and $A$ are the atomic and mass numbers of the nucleus.


\begin{thebibliography}{99}
\bibitem{glastsens}
See {\tt www-glast.slac.stanford.edu/software/IS/glast\_lat\_performance.htm}.
\end{thebibliography}
\end{document}